\documentclass[runningheads]{llncs}
\usepackage{enumitem}
\setlist[itemize]{noitemsep, nolistsep, topsep=0pt, partopsep={0pt}} 
\usepackage{float}
\usepackage{diagbox}
\usepackage{verbatim}
\newlist{Properties}{enumerate}{2}
\setlist[Properties]{label=Property \arabic*.,itemindent=*}
\usepackage{tikz}

\usepackage{tcolorbox} 

\usepackage[font=small,labelfont=bf]{caption}

\usepackage[T1]{fontenc}
\usepackage{graphicx}
\usepackage{mdframed}

\usepackage{prettyref}
\newcommand{\rref}[2][]{\prettyref{#2}}
\newrefformat{sec}{Sect.\,\ref{#1}}
\newrefformat{def}{Definition\,\ref{#1}}
\newrefformat{thm}{Theorem\,\ref{#1}}
\newrefformat{app}{Appendix\,\ref{#1}}
\newrefformat{prop}{Proposition\,\ref{#1}}
\newrefformat{rem}{Remark\,\ref{#1}}
\newrefformat{lem}{Lemma\,\ref{#1}}
\newrefformat{cor}{Corollary\,\ref{#1}}
\newrefformat{ex}{Example\,\ref{#1}}
\newrefformat{tab}{Table\,\ref{#1}}
\newrefformat{fig}{Fig.\,\ref{#1}}
\newrefformat{case}{case\,\ref{#1}}
\newrefformat{foot}{Footnote\,\ref{#1}}
\newrefformat{para}{Paragraph\,\ref{#1}}

\usepackage{xcolor}
\usepackage{colortbl}
\usepackage{isabelle}
\usepackage{isabellesym}
\usepackage{amsmath}
\usepackage{amssymb}
\usepackage{wrapfig}
\usepackage{subfig}
\usepackage{hyperref}

\newcommand{\snip}[4]{\expandafter\newcommand\csname #1\endcsname{#4}}
\snip{WESTandstatecorrect}{1}{2}{%
\isacommand{lemma}\isamarkupfalse%
\ WEST{\isacharunderscore}{\kern0pt}and{\isacharunderscore}{\kern0pt}state{\isacharunderscore}{\kern0pt}correct{\isacharcolon}{\kern0pt}\ \isanewline
\ \ \isakeyword{fixes}\ state{\isacharcolon}{\kern0pt}{\isacharcolon}{\kern0pt}{\isachardoublequoteopen}state{\isachardoublequoteclose}\ \isakeyword{and}\ s{\isadigit{1}}\ s{\isadigit{2}}{\isacharcolon}{\kern0pt}{\isacharcolon}{\kern0pt}\ {\isachardoublequoteopen}state{\isacharunderscore}{\kern0pt}regex{\isachardoublequoteclose}\isanewline
\ \ \isakeyword{assumes}\ {\isachardoublequoteopen}length\ s{\isadigit{1}}\ {\isacharequal}{\kern0pt}\ length\ s{\isadigit{2}}{\isachardoublequoteclose}\isanewline
\ \ \isakeyword{shows}\ {\isachardoublequoteopen}{\isacharparenleft}{\kern0pt}match{\isacharunderscore}{\kern0pt}timestep\ state\ s{\isadigit{1}}\ {\isasymand}\ match{\isacharunderscore}{\kern0pt}timestep\ state\ s{\isadigit{2}}{\isacharparenright}{\kern0pt}\ {\isasymlongleftrightarrow}\ \isanewline
\ \ \ \ \ \ \ \ \ {\isacharparenleft}{\kern0pt}{\isasymexists}somestate{\isachardot}{\kern0pt}\ match{\isacharunderscore}{\kern0pt}timestep\ state\ somestate\ {\isasymand}\ \isanewline
\ \ \ \ \ \ \ \ \ {\isacharparenleft}{\kern0pt}WEST{\isacharunderscore}{\kern0pt}and{\isacharunderscore}{\kern0pt}state\ s{\isadigit{1}}\ s{\isadigit{2}}{\isacharparenright}{\kern0pt}\ {\isacharequal}{\kern0pt}\ Some\ somestate{\isacharparenright}{\kern0pt}{\isachardoublequoteclose}%
}
\snip{WESTandcorrect}{1}{2}{%
\isacommand{lemma}\isamarkupfalse%
\ WEST{\isacharunderscore}{\kern0pt}and{\isacharunderscore}{\kern0pt}correct{\isacharcolon}{\kern0pt}\ \isanewline
\ \ \isakeyword{fixes}\ {\isasympi}{\isacharcolon}{\kern0pt}{\isacharcolon}{\kern0pt}{\isachardoublequoteopen}trace{\isachardoublequoteclose}\ \isakeyword{and}\ L{\isadigit{1}}\ L{\isadigit{2}}{\isacharcolon}{\kern0pt}{\isacharcolon}{\kern0pt}\ {\isachardoublequoteopen}WEST{\isacharunderscore}{\kern0pt}regex{\isachardoublequoteclose}\isanewline
\ \ \isakeyword{assumes}\ L{\isadigit{1}}{\isacharunderscore}{\kern0pt}of{\isacharunderscore}{\kern0pt}num{\isacharunderscore}{\kern0pt}vars{\isacharcolon}{\kern0pt}\ {\isachardoublequoteopen}WEST{\isacharunderscore}{\kern0pt}regex{\isacharunderscore}{\kern0pt}of{\isacharunderscore}{\kern0pt}vars\ L{\isadigit{1}}\ n{\isachardoublequoteclose}\isanewline
\ \ \isakeyword{assumes}\ L{\isadigit{2}}{\isacharunderscore}{\kern0pt}of{\isacharunderscore}{\kern0pt}num{\isacharunderscore}{\kern0pt}vars{\isacharcolon}{\kern0pt}\ {\isachardoublequoteopen}WEST{\isacharunderscore}{\kern0pt}regex{\isacharunderscore}{\kern0pt}of{\isacharunderscore}{\kern0pt}vars\ L{\isadigit{2}}\ n{\isachardoublequoteclose}\isanewline
\ \ \isakeyword{shows}\ {\isachardoublequoteopen}{\isacharparenleft}{\kern0pt}match\ {\isasympi}\ L{\isadigit{1}}\ {\isasymand}\ match\ {\isasympi}\ L{\isadigit{2}}{\isacharparenright}{\kern0pt}\ {\isasymlongleftrightarrow}\ match\ {\isasympi}\ {\isacharparenleft}{\kern0pt}WEST{\isacharunderscore}{\kern0pt}and\ L{\isadigit{1}}\ L{\isadigit{2}}{\isacharparenright}{\kern0pt}{\isachardoublequoteclose}%
}
\snip{WESTorcorrect}{1}{2}{%
\isacommand{lemma}\isamarkupfalse%
\ WEST{\isacharunderscore}{\kern0pt}or{\isacharunderscore}{\kern0pt}correct{\isacharcolon}{\kern0pt}\isanewline
\ \ \isakeyword{fixes}\ {\isasympi}{\isacharcolon}{\kern0pt}{\isacharcolon}{\kern0pt}{\isachardoublequoteopen}trace{\isachardoublequoteclose}\ \isakeyword{and}\ L{\isadigit{1}}\ L{\isadigit{2}}{\isacharcolon}{\kern0pt}{\isacharcolon}{\kern0pt}{\isachardoublequoteopen}WEST{\isacharunderscore}{\kern0pt}regex{\isachardoublequoteclose}\isanewline
\ \ \isakeyword{shows}\ {\isachardoublequoteopen}match\ {\isasympi}\ {\isacharparenleft}{\kern0pt}L{\isadigit{1}}{\isacharat}{\kern0pt}L{\isadigit{2}}{\isacharparenright}{\kern0pt}\ {\isasymlongleftrightarrow}\ {\isacharparenleft}{\kern0pt}match\ {\isasympi}\ L{\isadigit{1}}{\isacharparenright}{\kern0pt}\ {\isasymor}\ {\isacharparenleft}{\kern0pt}match\ {\isasympi}\ L{\isadigit{2}}{\isacharparenright}{\kern0pt}{\isachardoublequoteclose}%
}
\snip{WESTsimpcorrect}{1}{2}{%
\isacommand{lemma}\isamarkupfalse%
\ WEST{\isacharunderscore}{\kern0pt}simp{\isacharunderscore}{\kern0pt}correct{\isacharcolon}{\kern0pt}\ \isanewline
\ \ \isakeyword{fixes}\ L{\isacharcolon}{\kern0pt}{\isacharcolon}{\kern0pt}{\isachardoublequoteopen}WEST{\isacharunderscore}{\kern0pt}regex{\isachardoublequoteclose}\ \isakeyword{and}\ {\isasympi}{\isacharcolon}{\kern0pt}{\isacharcolon}{\kern0pt}{\isachardoublequoteopen}trace{\isachardoublequoteclose}\ \isakeyword{and}\ n{\isacharcolon}{\kern0pt}{\isacharcolon}{\kern0pt}{\isachardoublequoteopen}nat{\isachardoublequoteclose}\isanewline
\ \ \isakeyword{assumes}\ {\isachardoublequoteopen}WEST{\isacharunderscore}{\kern0pt}regex{\isacharunderscore}{\kern0pt}of{\isacharunderscore}{\kern0pt}vars\ L\ n{\isachardoublequoteclose}\isanewline
\ \ \isakeyword{shows}\ {\isachardoublequoteopen}match\ {\isasympi}\ {\isacharparenleft}{\kern0pt}WEST{\isacharunderscore}{\kern0pt}simp\ L\ n{\isacharparenright}{\kern0pt}\ {\isasymlongleftrightarrow}\ match\ {\isasympi}\ L{\isachardoublequoteclose}%
}
\snip{WESTglobalcorrect}{1}{2}{%
\isacommand{lemma}\isamarkupfalse%
\ WEST{\isacharunderscore}{\kern0pt}global{\isacharunderscore}{\kern0pt}correct{\isacharcolon}{\kern0pt}\ \isanewline
\isakeyword{fixes}\ L{\isacharcolon}{\kern0pt}{\isacharcolon}{\kern0pt}{\isachardoublequoteopen}WEST{\isacharunderscore}{\kern0pt}regex{\isachardoublequoteclose}\ \isakeyword{and}\ {\isasymphi}{\isacharcolon}{\kern0pt}{\isacharcolon}{\kern0pt}{\isachardoublequoteopen}nat\ mltl{\isachardoublequoteclose}\ \isakeyword{and}\ {\isasympi}{\isacharcolon}{\kern0pt}{\isacharcolon}{\kern0pt}{\isachardoublequoteopen}trace{\isachardoublequoteclose}\isanewline
\isakeyword{assumes}\ semantics{\isacharunderscore}{\kern0pt}{\isasymphi}{\isacharcolon}{\kern0pt}\ {\isachardoublequoteopen}{\isasymAnd}{\isasympi}{\isachardot}{\kern0pt}\ {\isacharparenleft}{\kern0pt}length\ {\isasympi}\ {\isasymge}\ complen{\isacharunderscore}{\kern0pt}mltl\ {\isasymphi}\ {\isasymlongrightarrow}\ \isanewline
\ \ \ \ \ \ \ \ \ \ \ \ \ \ \ \ \ \ \ \ {\isacharparenleft}{\kern0pt}match\ {\isasympi}\ L\ {\isasymlongleftrightarrow}\ semantics{\isacharunderscore}{\kern0pt}mltl\ {\isasympi}\ {\isasymphi}{\isacharparenright}{\kern0pt}{\isacharparenright}{\kern0pt}{\isachardoublequoteclose}\ \isanewline
\isakeyword{assumes}\ L{\isacharunderscore}{\kern0pt}vars{\isacharcolon}{\kern0pt}\ {\isachardoublequoteopen}WEST{\isacharunderscore}{\kern0pt}regex{\isacharunderscore}{\kern0pt}of{\isacharunderscore}{\kern0pt}vars\ L\ n{\isachardoublequoteclose}\isanewline
\isakeyword{assumes}\ {\isasymphi}{\isacharunderscore}{\kern0pt}vars{\isacharcolon}{\kern0pt}\ {\isachardoublequoteopen}WEST{\isacharunderscore}{\kern0pt}num{\isacharunderscore}{\kern0pt}vars\ {\isasymphi}\ {\isasymle}\ n{\isachardoublequoteclose}\ \isakeyword{and}\ {\isachardoublequoteopen}a{\isasymle}b{\isachardoublequoteclose}\isanewline
\isakeyword{assumes}\ trace{\isacharunderscore}{\kern0pt}len{\isacharcolon}{\kern0pt}\ {\isachardoublequoteopen}length\ {\isasympi}\ {\isasymge}\ {\isacharparenleft}{\kern0pt}complen{\isacharunderscore}{\kern0pt}mltl\ {\isasymphi}{\isacharparenright}{\kern0pt}\ {\isacharplus}{\kern0pt}\ b{\isachardoublequoteclose}\isanewline
\isakeyword{shows}\ {\isachardoublequoteopen}match\ {\isasympi}\ {\isacharparenleft}{\kern0pt}WEST{\isacharunderscore}{\kern0pt}global\ L\ a\ b\ n{\isacharparenright}{\kern0pt}\ {\isasymlongleftrightarrow}\ \isanewline
\ \ \ \ \ \ \ semantics{\isacharunderscore}{\kern0pt}mltl\ {\isasympi}\ {\isacharparenleft}{\kern0pt}Global{\isacharunderscore}{\kern0pt}mltl\ {\isasymphi}\ a\ b{\isacharparenright}{\kern0pt}{\isachardoublequoteclose}%
}
\snip{WESTregauxcorrect}{1}{2}{%
\isacommand{theorem}\isamarkupfalse%
\ WEST{\isacharunderscore}{\kern0pt}reg{\isacharunderscore}{\kern0pt}aux{\isacharunderscore}{\kern0pt}correct{\isacharcolon}{\kern0pt}\ \isanewline
\isakeyword{fixes}\ {\isasympi}{\isacharcolon}{\kern0pt}{\isacharcolon}{\kern0pt}{\isachardoublequoteopen}trace{\isachardoublequoteclose}\ \isakeyword{and}\ {\isasymphi}{\isacharcolon}{\kern0pt}{\isacharcolon}{\kern0pt}{\isachardoublequoteopen}nat\ mltl{\isachardoublequoteclose}\ \isakeyword{and}\ n{\isacharcolon}{\kern0pt}{\isacharcolon}{\kern0pt}{\isachardoublequoteopen}nat{\isachardoublequoteclose}\isanewline
\isakeyword{assumes}\ {\isasympi}{\isacharunderscore}{\kern0pt}long{\isacharunderscore}{\kern0pt}enough{\isacharcolon}{\kern0pt}\ {\isachardoublequoteopen}length\ {\isasympi}\ {\isasymge}\ complen{\isacharunderscore}{\kern0pt}mltl\ {\isasymphi}{\isachardoublequoteclose}\isanewline
\isakeyword{assumes}\ is{\isacharunderscore}{\kern0pt}nnf{\isacharcolon}{\kern0pt}\ {\isachardoublequoteopen}{\isasymexists}\ {\isasympsi}{\isachardot}{\kern0pt}\ {\isasymphi}\ {\isacharequal}{\kern0pt}\ {\isacharparenleft}{\kern0pt}convert{\isacharunderscore}{\kern0pt}nnf\ {\isasympsi}{\isacharparenright}{\kern0pt}{\isachardoublequoteclose}\isanewline
\isakeyword{assumes}\ {\isasymphi}{\isacharunderscore}{\kern0pt}nv{\isacharcolon}{\kern0pt}\ {\isachardoublequoteopen}WEST{\isacharunderscore}{\kern0pt}num{\isacharunderscore}{\kern0pt}vars\ {\isasymphi}\ {\isasymle}\ n{\isachardoublequoteclose}\isanewline
\isakeyword{assumes}\ {\isachardoublequoteopen}intervals{\isacharunderscore}{\kern0pt}welldef\ {\isasymphi}{\isachardoublequoteclose}\isanewline
\isakeyword{shows}\ {\isachardoublequoteopen}match\ {\isasympi}\ {\isacharparenleft}{\kern0pt}WEST{\isacharunderscore}{\kern0pt}reg{\isacharunderscore}{\kern0pt}aux\ {\isasymphi}\ n{\isacharparenright}{\kern0pt}\ {\isasymlongleftrightarrow}\ semantics{\isacharunderscore}{\kern0pt}mltl\ {\isasympi}\ {\isasymphi}{\isachardoublequoteclose}%
}
\snip{WESTregauxnv}{1}{2}{%
\isacommand{lemma}\isamarkupfalse%
\ WEST{\isacharunderscore}{\kern0pt}reg{\isacharunderscore}{\kern0pt}aux{\isacharunderscore}{\kern0pt}num{\isacharunderscore}{\kern0pt}vars{\isacharcolon}{\kern0pt}\ \isanewline
\isakeyword{fixes}\ {\isasymphi}{\isacharcolon}{\kern0pt}{\isacharcolon}{\kern0pt}{\isachardoublequoteopen}nat\ mltl{\isachardoublequoteclose}\isanewline
\isakeyword{assumes}\ is{\isacharunderscore}{\kern0pt}nnf{\isacharcolon}{\kern0pt}\ {\isachardoublequoteopen}{\isasymexists}\ {\isasympsi}{\isachardot}{\kern0pt}\ {\isasymphi}\ {\isacharequal}{\kern0pt}\ {\isacharparenleft}{\kern0pt}convert{\isacharunderscore}{\kern0pt}nnf\ {\isasympsi}{\isacharparenright}{\kern0pt}{\isachardoublequoteclose}\isanewline
\isakeyword{assumes}\ {\isachardoublequoteopen}WEST{\isacharunderscore}{\kern0pt}num{\isacharunderscore}{\kern0pt}vars\ {\isasymphi}\ {\isasymle}\ n{\isachardoublequoteclose}\ \isakeyword{and}\ {\isachardoublequoteopen}intervals{\isacharunderscore}{\kern0pt}welldef\ {\isasymphi}{\isachardoublequoteclose}\isanewline
\isakeyword{shows}\ {\isachardoublequoteopen}WEST{\isacharunderscore}{\kern0pt}regex{\isacharunderscore}{\kern0pt}of{\isacharunderscore}{\kern0pt}vars\ {\isacharparenleft}{\kern0pt}WEST{\isacharunderscore}{\kern0pt}reg{\isacharunderscore}{\kern0pt}aux\ {\isasymphi}\ n{\isacharparenright}{\kern0pt}\ n{\isachardoublequoteclose}%
}
\snip{WESTregcorrect}{1}{2}{%
\isacommand{theorem}\isamarkupfalse%
\ WEST{\isacharunderscore}{\kern0pt}reg{\isacharunderscore}{\kern0pt}correct{\isacharcolon}{\kern0pt}\ \isanewline
\ \ \isakeyword{fixes}\ {\isasymphi}{\isacharcolon}{\kern0pt}{\isacharcolon}{\kern0pt}{\isachardoublequoteopen}nat\ mltl{\isachardoublequoteclose}\ \isakeyword{and}\ {\isasympi}{\isacharcolon}{\kern0pt}{\isacharcolon}{\kern0pt}{\isachardoublequoteopen}trace{\isachardoublequoteclose}\isanewline
\ \ \isakeyword{assumes}\ {\isachardoublequoteopen}intervals{\isacharunderscore}{\kern0pt}welldef\ {\isasymphi}{\isachardoublequoteclose}\isanewline
\ \ \isakeyword{assumes}\ {\isasympi}{\isacharunderscore}{\kern0pt}long{\isacharunderscore}{\kern0pt}enough{\isacharcolon}{\kern0pt}\ {\isachardoublequoteopen}length\ {\isasympi}\ {\isasymge}\ complen{\isacharunderscore}{\kern0pt}mltl\ {\isasymphi}{\isachardoublequoteclose}\isanewline
\ \ \isakeyword{shows}\ \ {\isachardoublequoteopen}match\ {\isasympi}\ {\isacharparenleft}{\kern0pt}WEST{\isacharunderscore}{\kern0pt}reg\ {\isasymphi}{\isacharparenright}{\kern0pt}\ {\isasymlongleftrightarrow}\ semantics{\isacharunderscore}{\kern0pt}mltl\ {\isasympi}\ {\isasymphi}{\isachardoublequoteclose}%
}
\snip{regexequivalencecorrect}{1}{2}{%
\isacommand{lemma}\isamarkupfalse%
\ regex{\isacharunderscore}{\kern0pt}equivalence{\isacharunderscore}{\kern0pt}correct{\isacharcolon}{\kern0pt}\isanewline
\isakeyword{fixes}\ A\ B{\isacharcolon}{\kern0pt}{\isacharcolon}{\kern0pt}{\isachardoublequoteopen}WEST{\isacharunderscore}{\kern0pt}regex{\isachardoublequoteclose}\isanewline
\isakeyword{shows}\ {\isachardoublequoteopen}{\isacharparenleft}{\kern0pt}naive{\isacharunderscore}{\kern0pt}equivalence\ A\ B{\isacharparenright}{\kern0pt}\ {\isasymlongrightarrow}\ {\isacharparenleft}{\kern0pt}{\isasymforall}{\isasympi}{\isachardot}{\kern0pt}\ match\ {\isasympi}\ A\ {\isacharequal}{\kern0pt}\ match\ {\isasympi}\ B{\isacharparenright}{\kern0pt}{\isachardoublequoteclose}%
}
\snip{WESTsimphelper}{1}{2}{%
\isacommand{function}\isamarkupfalse%
\ WEST{\isacharunderscore}{\kern0pt}simp{\isacharunderscore}{\kern0pt}helper{\isacharcolon}{\kern0pt}{\isacharcolon}{\kern0pt}\ {\isachardoublequoteopen}WEST{\isacharunderscore}{\kern0pt}regex\ {\isasymRightarrow}\ {\isacharparenleft}{\kern0pt}nat\ {\isasymtimes}\ nat{\isacharparenright}{\kern0pt}\ list\ \isanewline
\ \ \ \ \ \ \ \ \ \ \ \ \ \ \ \ \ \ \ \ \ \ \ \ \ \ \ \ \ {\isasymRightarrow}\ nat\ {\isasymRightarrow}\ nat\ {\isasymRightarrow}\ WEST{\isacharunderscore}{\kern0pt}regex{\isachardoublequoteclose}\isanewline
\isakeyword{where}\ {\isachardoublequoteopen}WEST{\isacharunderscore}{\kern0pt}simp{\isacharunderscore}{\kern0pt}helper\ L\ idx{\isacharunderscore}{\kern0pt}pairs\ i\ n\ {\isacharequal}{\kern0pt}\ \isanewline
{\isacharparenleft}{\kern0pt}if\ {\isacharparenleft}{\kern0pt}idx{\isacharunderscore}{\kern0pt}pairs{\isasymnoteq}enum{\isacharunderscore}{\kern0pt}pairs\ L\ {\isasymor}\ i{\isasymge}length\ idx{\isacharunderscore}{\kern0pt}pairs{\isacharparenright}{\kern0pt}\ then\ L\ \isanewline
\ else\ {\isacharparenleft}{\kern0pt}if\ {\isacharparenleft}{\kern0pt}check{\isacharunderscore}{\kern0pt}simp\ {\isacharparenleft}{\kern0pt}L{\isacharbang}{\kern0pt}{\isacharparenleft}{\kern0pt}fst\ {\isacharparenleft}{\kern0pt}idx{\isacharunderscore}{\kern0pt}pairs{\isacharbang}{\kern0pt}i{\isacharparenright}{\kern0pt}{\isacharparenright}{\kern0pt}{\isacharparenright}{\kern0pt}\ \isanewline
\ \ \ \ \ \ \ \ \ \ \ \ \ \ \ \ \ \ \ \ \ \ {\isacharparenleft}{\kern0pt}L{\isacharbang}{\kern0pt}{\isacharparenleft}{\kern0pt}snd\ {\isacharparenleft}{\kern0pt}idx{\isacharunderscore}{\kern0pt}pairs{\isacharbang}{\kern0pt}i{\isacharparenright}{\kern0pt}{\isacharparenright}{\kern0pt}{\isacharparenright}{\kern0pt}{\isacharparenright}{\kern0pt}\ then\isanewline
\ \ \ \ {\isacharparenleft}{\kern0pt}let\ newL\ {\isacharequal}{\kern0pt}\ update{\isacharunderscore}{\kern0pt}L\ L\ {\isacharparenleft}{\kern0pt}idx{\isacharunderscore}{\kern0pt}pairs{\isacharbang}{\kern0pt}i{\isacharparenright}{\kern0pt}\ n\ in\ \isanewline
\ \ \ \ WEST{\isacharunderscore}{\kern0pt}simp{\isacharunderscore}{\kern0pt}helper\ newL\ {\isacharparenleft}{\kern0pt}enum{\isacharunderscore}{\kern0pt}pairs\ newL{\isacharparenright}{\kern0pt}\ {\isadigit{0}}\ n{\isacharparenright}{\kern0pt}\ \isanewline
\ \ \ \ else\ WEST{\isacharunderscore}{\kern0pt}simp{\isacharunderscore}{\kern0pt}helper\ L\ idx{\isacharunderscore}{\kern0pt}pairs\ {\isacharparenleft}{\kern0pt}i{\isacharplus}{\kern0pt}{\isadigit{1}}{\isacharparenright}{\kern0pt}\ n{\isacharparenright}{\kern0pt}{\isacharparenright}{\kern0pt}{\isachardoublequoteclose}%
}
\snip{WESTsimp}{1}{2}{%
\isacommand{fun}\isamarkupfalse%
\ WEST{\isacharunderscore}{\kern0pt}simp{\isacharcolon}{\kern0pt}{\isacharcolon}{\kern0pt}\ {\isachardoublequoteopen}WEST{\isacharunderscore}{\kern0pt}regex\ {\isasymRightarrow}\ nat\ {\isasymRightarrow}\ WEST{\isacharunderscore}{\kern0pt}regex{\isachardoublequoteclose}\isanewline
\ \ \isakeyword{where}\ {\isachardoublequoteopen}WEST{\isacharunderscore}{\kern0pt}simp\ L\ num{\isacharunderscore}{\kern0pt}vars\ {\isacharequal}{\kern0pt}\ \isanewline
\ \ WEST{\isacharunderscore}{\kern0pt}simp{\isacharunderscore}{\kern0pt}helper\ L\ {\isacharparenleft}{\kern0pt}enum{\isacharunderscore}{\kern0pt}pairs\ L{\isacharparenright}{\kern0pt}\ {\isadigit{0}}\ num{\isacharunderscore}{\kern0pt}vars{\isachardoublequoteclose}%
}
\snip{WESTandsimp}{1}{2}{%
\isacommand{fun}\isamarkupfalse%
\ WEST{\isacharunderscore}{\kern0pt}and{\isacharunderscore}{\kern0pt}simp{\isacharcolon}{\kern0pt}{\isacharcolon}{\kern0pt}\ {\isachardoublequoteopen}WEST{\isacharunderscore}{\kern0pt}regex\ {\isasymRightarrow}\ WEST{\isacharunderscore}{\kern0pt}regex\ \isanewline
\ \ \ \ \ \ \ \ \ \ \ \ \ \ \ \ \ \ \ \ \ {\isasymRightarrow}\ nat\ {\isasymRightarrow}\ WEST{\isacharunderscore}{\kern0pt}regex{\isachardoublequoteclose}\isanewline
\ \ \isakeyword{where}\ {\isachardoublequoteopen}WEST{\isacharunderscore}{\kern0pt}and{\isacharunderscore}{\kern0pt}simp\ L{\isadigit{1}}\ L{\isadigit{2}}\ num{\isacharunderscore}{\kern0pt}vars\ {\isacharequal}{\kern0pt}\ \isanewline
\ \ \ \ \ \ \ \ \ WEST{\isacharunderscore}{\kern0pt}simp\ {\isacharparenleft}{\kern0pt}WEST{\isacharunderscore}{\kern0pt}and\ L{\isadigit{1}}\ L{\isadigit{2}}{\isacharparenright}{\kern0pt}\ num{\isacharunderscore}{\kern0pt}vars{\isachardoublequoteclose}%
}
\snip{WESTorsimp}{1}{2}{%
\isacommand{fun}\isamarkupfalse%
\ WEST{\isacharunderscore}{\kern0pt}or{\isacharunderscore}{\kern0pt}simp{\isacharcolon}{\kern0pt}{\isacharcolon}{\kern0pt}\ {\isachardoublequoteopen}WEST{\isacharunderscore}{\kern0pt}regex\ {\isasymRightarrow}\ WEST{\isacharunderscore}{\kern0pt}regex\ {\isasymRightarrow}\ nat\ {\isasymRightarrow}\ WEST{\isacharunderscore}{\kern0pt}regex{\isachardoublequoteclose}\isanewline
\isakeyword{where}\ {\isachardoublequoteopen}WEST{\isacharunderscore}{\kern0pt}or{\isacharunderscore}{\kern0pt}simp\ L{\isadigit{1}}\ L{\isadigit{2}}\ num{\isacharunderscore}{\kern0pt}vars\ {\isacharequal}{\kern0pt}\ WEST{\isacharunderscore}{\kern0pt}simp\ {\isacharparenleft}{\kern0pt}L{\isadigit{1}}{\isacharat}{\kern0pt}L{\isadigit{2}}{\isacharparenright}{\kern0pt}\ num{\isacharunderscore}{\kern0pt}vars{\isachardoublequoteclose}%
}
\snip{padmatchproperty}{1}{2}{%
\isacommand{lemma}\isamarkupfalse%
\ shift{\isacharunderscore}{\kern0pt}match{\isacharunderscore}{\kern0pt}property{\isacharcolon}{\kern0pt}\isanewline
\ \ \isakeyword{assumes}\ {\isachardoublequoteopen}length\ {\isasympi}\ {\isasymge}\ t{\isachardoublequoteclose}\isanewline
\ \ \isakeyword{shows}\ {\isachardoublequoteopen}match\ {\isacharparenleft}{\kern0pt}drop\ t\ {\isasympi}{\isacharparenright}{\kern0pt}\ L\ {\isasymlongleftrightarrow}\ match\ {\isasympi}\ {\isacharparenleft}{\kern0pt}shift\ L\ num{\isacharunderscore}{\kern0pt}vars\ t{\isacharparenright}{\kern0pt}{\isachardoublequoteclose}%
}
\snip{arbitrarystate}{1}{2}{%
\isacommand{fun}\isamarkupfalse%
\ arbitrary{\isacharunderscore}{\kern0pt}state{\isacharcolon}{\kern0pt}{\isacharcolon}{\kern0pt}{\isachardoublequoteopen}nat\ {\isasymRightarrow}\ state{\isacharunderscore}{\kern0pt}regex{\isachardoublequoteclose}\isanewline
\ \ \isakeyword{where}\ {\isachardoublequoteopen}arbitrary{\isacharunderscore}{\kern0pt}state\ num{\isacharunderscore}{\kern0pt}vars\ {\isacharequal}{\kern0pt}\ map\ {\isacharparenleft}{\kern0pt}{\isasymlambda}\ k{\isachardot}{\kern0pt}\ S{\isacharparenright}{\kern0pt}\ {\isacharbrackleft}{\kern0pt}{\isadigit{0}}\ {\isachardot}{\kern0pt}{\isachardot}{\kern0pt}{\isacharless}{\kern0pt}\ num{\isacharunderscore}{\kern0pt}vars{\isacharbrackright}{\kern0pt}{\isachardoublequoteclose}%
}
\snip{arbitrarytrace}{1}{2}{%
\isacommand{fun}\isamarkupfalse%
\ arbitrary{\isacharunderscore}{\kern0pt}trace{\isacharcolon}{\kern0pt}{\isacharcolon}{\kern0pt}{\isachardoublequoteopen}nat\ {\isasymRightarrow}\ nat\ {\isasymRightarrow}\ trace{\isacharunderscore}{\kern0pt}regex{\isachardoublequoteclose}\isanewline
\ \ \isakeyword{where}\ {\isachardoublequoteopen}arbitrary{\isacharunderscore}{\kern0pt}trace\ num{\isacharunderscore}{\kern0pt}vars\ num{\isacharunderscore}{\kern0pt}shift\ {\isacharequal}{\kern0pt}\ \isanewline
\ \ map\ {\isacharparenleft}{\kern0pt}{\isasymlambda}\ k{\isachardot}{\kern0pt}\ {\isacharparenleft}{\kern0pt}arbitrary{\isacharunderscore}{\kern0pt}state\ num{\isacharunderscore}{\kern0pt}vars{\isacharparenright}{\kern0pt}{\isacharparenright}{\kern0pt}\ {\isacharbrackleft}{\kern0pt}{\isadigit{0}}\ {\isachardot}{\kern0pt}{\isachardot}{\kern0pt}{\isacharless}{\kern0pt}\ num{\isacharunderscore}{\kern0pt}shift{\isacharbrackright}{\kern0pt}{\isachardoublequoteclose}%
}
\snip{shift}{1}{2}{%
\isacommand{fun}\isamarkupfalse%
\ shift{\isacharcolon}{\kern0pt}{\isacharcolon}{\kern0pt}\ {\isachardoublequoteopen}WEST{\isacharunderscore}{\kern0pt}regex\ {\isasymRightarrow}\ nat\ {\isasymRightarrow}\ nat\ {\isasymRightarrow}\ WEST{\isacharunderscore}{\kern0pt}regex{\isachardoublequoteclose}\isanewline
\isakeyword{where}\ {\isachardoublequoteopen}shift\ L\ n\ t\ {\isacharequal}{\kern0pt}\ map\ {\isacharparenleft}{\kern0pt}{\isasymlambda}trace{\isachardot}{\kern0pt}\ {\isacharparenleft}{\kern0pt}arbitrary{\isacharunderscore}{\kern0pt}trace\ n\ t{\isacharparenright}{\kern0pt}{\isacharat}{\kern0pt}trace{\isacharparenright}{\kern0pt}\ L{\isachardoublequoteclose}%
}
\snip{WESTglobal}{1}{2}{%
\isacommand{fun}\isamarkupfalse%
\ WEST{\isacharunderscore}{\kern0pt}global{\isacharcolon}{\kern0pt}{\isacharcolon}{\kern0pt}\ {\isachardoublequoteopen}WEST{\isacharunderscore}{\kern0pt}regex\ {\isasymRightarrow}\ nat\ {\isasymRightarrow}\ nat\ {\isasymRightarrow}\ nat\ {\isasymRightarrow}\ WEST{\isacharunderscore}{\kern0pt}regex{\isachardoublequoteclose}\ \isanewline
\isakeyword{where}\ {\isachardoublequoteopen}WEST{\isacharunderscore}{\kern0pt}global\ L\ a\ b\ n\ {\isacharequal}{\kern0pt}\ {\isacharparenleft}{\kern0pt}if\ {\isacharparenleft}{\kern0pt}a\ {\isacharequal}{\kern0pt}\ b{\isacharparenright}{\kern0pt}\ then\ {\isacharparenleft}{\kern0pt}shift\ L\ n\ a{\isacharparenright}{\kern0pt}\isanewline
else\ {\isacharparenleft}{\kern0pt}if\ {\isacharparenleft}{\kern0pt}a\ {\isacharless}{\kern0pt}\ b{\isacharparenright}{\kern0pt}\ then\ {\isacharparenleft}{\kern0pt}WEST{\isacharunderscore}{\kern0pt}and{\isacharunderscore}{\kern0pt}simp\ {\isacharparenleft}{\kern0pt}shift\ L\ n\ b{\isacharparenright}{\kern0pt}\ \isanewline
\ \ \ \ \ \ \ \ \ \ \ \ \ \ \ \ {\isacharparenleft}{\kern0pt}WEST{\isacharunderscore}{\kern0pt}global\ L\ a\ {\isacharparenleft}{\kern0pt}b{\isacharminus}{\kern0pt}{\isadigit{1}}{\isacharparenright}{\kern0pt}\ n{\isacharparenright}{\kern0pt}\ n{\isacharparenright}{\kern0pt}\ else\ {\isacharbrackleft}{\kern0pt}{\isacharbrackright}{\kern0pt}{\isacharparenright}{\kern0pt}{\isacharparenright}{\kern0pt}{\isachardoublequoteclose}%
}
\snip{WESTrelease_helper}{1}{2}{%
\isacommand{fun}\isamarkupfalse%
\ WEST{\isacharunderscore}{\kern0pt}release{\isacharunderscore}{\kern0pt}helper{\isacharcolon}{\kern0pt}{\isacharcolon}{\kern0pt}\ \isanewline
{\isachardoublequoteopen}WEST{\isacharunderscore}{\kern0pt}regex\ {\isasymRightarrow}\ WEST{\isacharunderscore}{\kern0pt}regex\ {\isasymRightarrow}\ nat\ {\isasymRightarrow}\ nat\ {\isasymRightarrow}\ nat\ {\isasymRightarrow}\ WEST{\isacharunderscore}{\kern0pt}regex{\isachardoublequoteclose}\ \isanewline
\ \ \isakeyword{where}\ {\isachardoublequoteopen}WEST{\isacharunderscore}{\kern0pt}release{\isacharunderscore}{\kern0pt}helper\ L{\isacharunderscore}{\kern0pt}{\isasymphi}\ L{\isacharunderscore}{\kern0pt}{\isasympsi}\ a\ ub\ num{\isacharunderscore}{\kern0pt}vars\ {\isacharequal}{\kern0pt}\ {\isacharparenleft}{\kern0pt}\isanewline
\ \ if\ {\isacharparenleft}{\kern0pt}a{\isacharequal}{\kern0pt}ub{\isacharparenright}{\kern0pt}\ then\ {\isacharparenleft}{\kern0pt}WEST{\isacharunderscore}{\kern0pt}and{\isacharunderscore}{\kern0pt}simp\ {\isacharparenleft}{\kern0pt}WEST{\isacharunderscore}{\kern0pt}global\ L{\isacharunderscore}{\kern0pt}{\isasymphi}\ a\ a\ num{\isacharunderscore}{\kern0pt}vars{\isacharparenright}{\kern0pt}\ \isanewline
\ \ \ \ \ \ \ \ \ \ \ \ \ \ \ \ \ \ \ \ \ \ {\isacharparenleft}{\kern0pt}WEST{\isacharunderscore}{\kern0pt}global\ L{\isacharunderscore}{\kern0pt}{\isasympsi}\ a\ a\ num{\isacharunderscore}{\kern0pt}vars{\isacharparenright}{\kern0pt}\ num{\isacharunderscore}{\kern0pt}vars{\isacharparenright}{\kern0pt}\isanewline
\ \ else\ {\isacharparenleft}{\kern0pt}if\ {\isacharparenleft}{\kern0pt}a\ {\isacharless}{\kern0pt}\ ub{\isacharparenright}{\kern0pt}\ then\ \isanewline
\ \ \ \ \ WEST{\isacharunderscore}{\kern0pt}or{\isacharunderscore}{\kern0pt}simp\ {\isacharparenleft}{\kern0pt}WEST{\isacharunderscore}{\kern0pt}release{\isacharunderscore}{\kern0pt}helper\ L{\isacharunderscore}{\kern0pt}{\isasymphi}\ L{\isacharunderscore}{\kern0pt}{\isasympsi}\ a\ {\isacharparenleft}{\kern0pt}ub{\isacharminus}{\kern0pt}{\isadigit{1}}{\isacharparenright}{\kern0pt}\ num{\isacharunderscore}{\kern0pt}vars{\isacharparenright}{\kern0pt}\isanewline
\ \ \ \ \ \ \ \ \ \ {\isacharparenleft}{\kern0pt}WEST{\isacharunderscore}{\kern0pt}and{\isacharunderscore}{\kern0pt}simp\ {\isacharparenleft}{\kern0pt}WEST{\isacharunderscore}{\kern0pt}global\ L{\isacharunderscore}{\kern0pt}{\isasympsi}\ a\ ub\ num{\isacharunderscore}{\kern0pt}vars{\isacharparenright}{\kern0pt}\ \isanewline
\ \ \ \ \ \ \ \ \ \ {\isacharparenleft}{\kern0pt}WEST{\isacharunderscore}{\kern0pt}global\ L{\isacharunderscore}{\kern0pt}{\isasymphi}\ ub\ ub\ num{\isacharunderscore}{\kern0pt}vars{\isacharparenright}{\kern0pt}\ num{\isacharunderscore}{\kern0pt}vars{\isacharparenright}{\kern0pt}\ num{\isacharunderscore}{\kern0pt}vars\isanewline
\ \ \ \ else\ {\isacharbrackleft}{\kern0pt}{\isacharbrackright}{\kern0pt}{\isacharparenright}{\kern0pt}{\isacharparenright}{\kern0pt}{\isachardoublequoteclose}%
}
\snip{WESTrelease}{1}{2}{%
\isacommand{fun}\isamarkupfalse%
\ WEST{\isacharunderscore}{\kern0pt}release{\isacharcolon}{\kern0pt}{\isacharcolon}{\kern0pt}\ {\isachardoublequoteopen}WEST{\isacharunderscore}{\kern0pt}regex\ {\isasymRightarrow}\ WEST{\isacharunderscore}{\kern0pt}regex\ {\isasymRightarrow}\ \ \isanewline
\ \ \ \ \ \ \ \ \ \ \ \ \ \ \ nat\ {\isasymRightarrow}\ nat\ {\isasymRightarrow}\ nat\ {\isasymRightarrow}\ WEST{\isacharunderscore}{\kern0pt}regex{\isachardoublequoteclose}\ \isanewline
\ \ \isakeyword{where}\ {\isachardoublequoteopen}WEST{\isacharunderscore}{\kern0pt}release\ L{\isacharunderscore}{\kern0pt}{\isasymphi}\ L{\isacharunderscore}{\kern0pt}{\isasympsi}\ a\ b\ num{\isacharunderscore}{\kern0pt}vars\ {\isacharequal}{\kern0pt}\ {\isacharparenleft}{\kern0pt}\isanewline
\ \ if\ {\isacharparenleft}{\kern0pt}b\ {\isachargreater}{\kern0pt}\ a{\isacharparenright}{\kern0pt}\ then\ {\isacharparenleft}{\kern0pt}WEST{\isacharunderscore}{\kern0pt}or{\isacharunderscore}{\kern0pt}simp\ {\isacharparenleft}{\kern0pt}WEST{\isacharunderscore}{\kern0pt}global\ L{\isacharunderscore}{\kern0pt}{\isasympsi}\ a\ b\ num{\isacharunderscore}{\kern0pt}vars{\isacharparenright}{\kern0pt}\ \isanewline
\ \ \ \ \ {\isacharparenleft}{\kern0pt}WEST{\isacharunderscore}{\kern0pt}release{\isacharunderscore}{\kern0pt}helper\ L{\isacharunderscore}{\kern0pt}{\isasymphi}\ L{\isacharunderscore}{\kern0pt}{\isasympsi}\ a\ {\isacharparenleft}{\kern0pt}b{\isacharminus}{\kern0pt}{\isadigit{1}}{\isacharparenright}{\kern0pt}\ num{\isacharunderscore}{\kern0pt}vars{\isacharparenright}{\kern0pt}\ num{\isacharunderscore}{\kern0pt}vars{\isacharparenright}{\kern0pt}\isanewline
\ \ else\ {\isacharparenleft}{\kern0pt}WEST{\isacharunderscore}{\kern0pt}global\ L{\isacharunderscore}{\kern0pt}{\isasympsi}\ a\ b\ num{\isacharunderscore}{\kern0pt}vars{\isacharparenright}{\kern0pt}{\isacharparenright}{\kern0pt}{\isachardoublequoteclose}%
}
\snip{WESTnumvars}{1}{2}{%
\isacommand{fun}\isamarkupfalse%
\ WEST{\isacharunderscore}{\kern0pt}num{\isacharunderscore}{\kern0pt}vars{\isacharcolon}{\kern0pt}{\isacharcolon}{\kern0pt}\ {\isachardoublequoteopen}{\isacharparenleft}{\kern0pt}nat{\isacharparenright}{\kern0pt}\ mltl\ {\isasymRightarrow}\ nat{\isachardoublequoteclose}\isanewline
\ \ \isakeyword{where}\ {\isachardoublequoteopen}WEST{\isacharunderscore}{\kern0pt}num{\isacharunderscore}{\kern0pt}vars\ True{\isacharunderscore}{\kern0pt}mltl\ {\isacharequal}{\kern0pt}\ {\isadigit{1}}{\isachardoublequoteclose}\isanewline
\ \ {\isacharbar}{\kern0pt}\ {\isachardoublequoteopen}WEST{\isacharunderscore}{\kern0pt}num{\isacharunderscore}{\kern0pt}vars\ False{\isacharunderscore}{\kern0pt}mltl\ {\isacharequal}{\kern0pt}\ {\isadigit{1}}{\isachardoublequoteclose}\isanewline
\ \ {\isacharbar}{\kern0pt}\ {\isachardoublequoteopen}WEST{\isacharunderscore}{\kern0pt}num{\isacharunderscore}{\kern0pt}vars\ {\isacharparenleft}{\kern0pt}Prop{\isacharunderscore}{\kern0pt}mltl\ p{\isacharparenright}{\kern0pt}\ {\isacharequal}{\kern0pt}\ p{\isacharplus}{\kern0pt}{\isadigit{1}}{\isachardoublequoteclose}\isanewline
\ \ {\isacharbar}{\kern0pt}\ {\isachardoublequoteopen}WEST{\isacharunderscore}{\kern0pt}num{\isacharunderscore}{\kern0pt}vars\ {\isacharparenleft}{\kern0pt}Not{\isacharunderscore}{\kern0pt}mltl\ {\isasymphi}{\isacharparenright}{\kern0pt}\ {\isacharequal}{\kern0pt}\ WEST{\isacharunderscore}{\kern0pt}num{\isacharunderscore}{\kern0pt}vars\ {\isasymphi}{\isachardoublequoteclose}\isanewline
\ \ {\isacharbar}{\kern0pt}\ {\isachardoublequoteopen}WEST{\isacharunderscore}{\kern0pt}num{\isacharunderscore}{\kern0pt}vars\ {\isacharparenleft}{\kern0pt}And{\isacharunderscore}{\kern0pt}mltl\ {\isasymphi}\ {\isasympsi}{\isacharparenright}{\kern0pt}\ {\isacharequal}{\kern0pt}\ \isanewline
\ \ \ \ \ Max\ {\isacharbraceleft}{\kern0pt}{\isacharparenleft}{\kern0pt}WEST{\isacharunderscore}{\kern0pt}num{\isacharunderscore}{\kern0pt}vars\ {\isasymphi}{\isacharparenright}{\kern0pt}{\isacharcomma}{\kern0pt}\ {\isacharparenleft}{\kern0pt}WEST{\isacharunderscore}{\kern0pt}num{\isacharunderscore}{\kern0pt}vars\ {\isasympsi}{\isacharparenright}{\kern0pt}{\isacharbraceright}{\kern0pt}{\isachardoublequoteclose}\isanewline
\ \ {\isacharbar}{\kern0pt}\ {\isachardoublequoteopen}WEST{\isacharunderscore}{\kern0pt}num{\isacharunderscore}{\kern0pt}vars\ {\isacharparenleft}{\kern0pt}Or{\isacharunderscore}{\kern0pt}mltl\ {\isasymphi}\ {\isasympsi}{\isacharparenright}{\kern0pt}\ {\isacharequal}{\kern0pt}\ \isanewline
\ \ \ \ \ Max\ {\isacharbraceleft}{\kern0pt}{\isacharparenleft}{\kern0pt}WEST{\isacharunderscore}{\kern0pt}num{\isacharunderscore}{\kern0pt}vars\ {\isasymphi}{\isacharparenright}{\kern0pt}{\isacharcomma}{\kern0pt}\ {\isacharparenleft}{\kern0pt}WEST{\isacharunderscore}{\kern0pt}num{\isacharunderscore}{\kern0pt}vars\ {\isasympsi}{\isacharparenright}{\kern0pt}{\isacharbraceright}{\kern0pt}{\isachardoublequoteclose}\isanewline
\ \ {\isacharbar}{\kern0pt}\ {\isachardoublequoteopen}WEST{\isacharunderscore}{\kern0pt}num{\isacharunderscore}{\kern0pt}vars\ {\isacharparenleft}{\kern0pt}Future{\isacharunderscore}{\kern0pt}mltl\ {\isasymphi}\ a\ b{\isacharparenright}{\kern0pt}\ {\isacharequal}{\kern0pt}\ WEST{\isacharunderscore}{\kern0pt}num{\isacharunderscore}{\kern0pt}vars\ {\isasymphi}{\isachardoublequoteclose}\isanewline
\ \ {\isacharbar}{\kern0pt}\ {\isachardoublequoteopen}WEST{\isacharunderscore}{\kern0pt}num{\isacharunderscore}{\kern0pt}vars\ {\isacharparenleft}{\kern0pt}Global{\isacharunderscore}{\kern0pt}mltl\ {\isasymphi}\ a\ b{\isacharparenright}{\kern0pt}\ {\isacharequal}{\kern0pt}\ WEST{\isacharunderscore}{\kern0pt}num{\isacharunderscore}{\kern0pt}vars\ {\isasymphi}{\isachardoublequoteclose}\isanewline
\ \ {\isacharbar}{\kern0pt}\ {\isachardoublequoteopen}WEST{\isacharunderscore}{\kern0pt}num{\isacharunderscore}{\kern0pt}vars\ {\isacharparenleft}{\kern0pt}Until{\isacharunderscore}{\kern0pt}mltl\ {\isasymphi}\ {\isasympsi}\ a\ b{\isacharparenright}{\kern0pt}\ {\isacharequal}{\kern0pt}\ \isanewline
\ \ \ \ \ Max\ {\isacharbraceleft}{\kern0pt}{\isacharparenleft}{\kern0pt}WEST{\isacharunderscore}{\kern0pt}num{\isacharunderscore}{\kern0pt}vars\ {\isasymphi}{\isacharparenright}{\kern0pt}{\isacharcomma}{\kern0pt}\ {\isacharparenleft}{\kern0pt}WEST{\isacharunderscore}{\kern0pt}num{\isacharunderscore}{\kern0pt}vars\ {\isasympsi}{\isacharparenright}{\kern0pt}{\isacharbraceright}{\kern0pt}{\isachardoublequoteclose}\isanewline
\ \ {\isacharbar}{\kern0pt}\ {\isachardoublequoteopen}WEST{\isacharunderscore}{\kern0pt}num{\isacharunderscore}{\kern0pt}vars\ {\isacharparenleft}{\kern0pt}Release{\isacharunderscore}{\kern0pt}mltl\ {\isasymphi}\ {\isasympsi}\ a\ b{\isacharparenright}{\kern0pt}\ {\isacharequal}{\kern0pt}\ \isanewline
\ \ \ \ \ Max\ {\isacharbraceleft}{\kern0pt}{\isacharparenleft}{\kern0pt}WEST{\isacharunderscore}{\kern0pt}num{\isacharunderscore}{\kern0pt}vars\ {\isasymphi}{\isacharparenright}{\kern0pt}{\isacharcomma}{\kern0pt}\ {\isacharparenleft}{\kern0pt}WEST{\isacharunderscore}{\kern0pt}num{\isacharunderscore}{\kern0pt}vars\ {\isasympsi}{\isacharparenright}{\kern0pt}{\isacharbraceright}{\kern0pt}{\isachardoublequoteclose}%
}
\snip{WESTregaux}{1}{2}{%
\isacommand{function}\isamarkupfalse%
\ WEST{\isacharunderscore}{\kern0pt}reg{\isacharunderscore}{\kern0pt}aux{\isacharcolon}{\kern0pt}{\isacharcolon}{\kern0pt}\ {\isachardoublequoteopen}nat\ mltl\ {\isasymRightarrow}\ nat\ {\isasymRightarrow}\ WEST{\isacharunderscore}{\kern0pt}regex{\isachardoublequoteclose}\ \isanewline
\isakeyword{where}\ {\isachardoublequoteopen}WEST{\isacharunderscore}{\kern0pt}reg{\isacharunderscore}{\kern0pt}aux\ True{\isacharunderscore}{\kern0pt}mltl\ n\ {\isacharequal}{\kern0pt}\ {\isacharbrackleft}{\kern0pt}{\isacharbrackleft}{\kern0pt}{\isacharparenleft}{\kern0pt}map\ {\isacharparenleft}{\kern0pt}{\isasymlambda}\ j{\isachardot}{\kern0pt}\ S{\isacharparenright}{\kern0pt}\ {\isacharbrackleft}{\kern0pt}{\isadigit{0}}\ {\isachardot}{\kern0pt}{\isachardot}{\kern0pt}{\isacharless}{\kern0pt}\ n{\isacharbrackright}{\kern0pt}{\isacharparenright}{\kern0pt}{\isacharbrackright}{\kern0pt}{\isacharbrackright}{\kern0pt}{\isachardoublequoteclose}\isanewline
{\isacharbar}{\kern0pt}\ {\isachardoublequoteopen}WEST{\isacharunderscore}{\kern0pt}reg{\isacharunderscore}{\kern0pt}aux\ False{\isacharunderscore}{\kern0pt}mltl\ n\ {\isacharequal}{\kern0pt}\ {\isacharbrackleft}{\kern0pt}{\isacharbrackright}{\kern0pt}{\isachardoublequoteclose}\isanewline
{\isacharbar}{\kern0pt}\ {\isachardoublequoteopen}WEST{\isacharunderscore}{\kern0pt}reg{\isacharunderscore}{\kern0pt}aux\ {\isacharparenleft}{\kern0pt}Prop{\isacharunderscore}{\kern0pt}mltl\ p{\isacharparenright}{\kern0pt}\ n\ {\isacharequal}{\kern0pt}\ \isanewline
\ \ \ {\isacharbrackleft}{\kern0pt}{\isacharbrackleft}{\kern0pt}{\isacharparenleft}{\kern0pt}map\ {\isacharparenleft}{\kern0pt}{\isasymlambda}j{\isachardot}{\kern0pt}\ {\isacharparenleft}{\kern0pt}if\ {\isacharparenleft}{\kern0pt}p{\isacharequal}{\kern0pt}j{\isacharparenright}{\kern0pt}\ then\ One\ else\ S{\isacharparenright}{\kern0pt}{\isacharparenright}{\kern0pt}\ {\isacharbrackleft}{\kern0pt}{\isadigit{0}}\ {\isachardot}{\kern0pt}{\isachardot}{\kern0pt}{\isacharless}{\kern0pt}\ n{\isacharbrackright}{\kern0pt}{\isacharparenright}{\kern0pt}{\isacharbrackright}{\kern0pt}{\isacharbrackright}{\kern0pt}{\isachardoublequoteclose}\isanewline
{\isacharbar}{\kern0pt}\ {\isachardoublequoteopen}WEST{\isacharunderscore}{\kern0pt}reg{\isacharunderscore}{\kern0pt}aux\ {\isacharparenleft}{\kern0pt}Not{\isacharunderscore}{\kern0pt}mltl\ {\isacharparenleft}{\kern0pt}Prop{\isacharunderscore}{\kern0pt}mltl\ p{\isacharparenright}{\kern0pt}{\isacharparenright}{\kern0pt}\ n\ {\isacharequal}{\kern0pt}\ \isanewline
\ \ \ {\isacharbrackleft}{\kern0pt}{\isacharbrackleft}{\kern0pt}{\isacharparenleft}{\kern0pt}map\ {\isacharparenleft}{\kern0pt}{\isasymlambda}j{\isachardot}{\kern0pt}\ {\isacharparenleft}{\kern0pt}if\ {\isacharparenleft}{\kern0pt}p{\isacharequal}{\kern0pt}j{\isacharparenright}{\kern0pt}\ then\ Zero\ else\ S{\isacharparenright}{\kern0pt}{\isacharparenright}{\kern0pt}\ {\isacharbrackleft}{\kern0pt}{\isadigit{0}}\ {\isachardot}{\kern0pt}{\isachardot}{\kern0pt}{\isacharless}{\kern0pt}\ n{\isacharbrackright}{\kern0pt}{\isacharparenright}{\kern0pt}{\isacharbrackright}{\kern0pt}{\isacharbrackright}{\kern0pt}{\isachardoublequoteclose}\isanewline
{\isacharbar}{\kern0pt}\ {\isachardoublequoteopen}WEST{\isacharunderscore}{\kern0pt}reg{\isacharunderscore}{\kern0pt}aux\ {\isacharparenleft}{\kern0pt}Or{\isacharunderscore}{\kern0pt}mltl\ {\isasymphi}\ {\isasympsi}{\isacharparenright}{\kern0pt}\ n\ {\isacharequal}{\kern0pt}\ WEST{\isacharunderscore}{\kern0pt}or{\isacharunderscore}{\kern0pt}simp\ \isanewline
\ \ \ {\isacharparenleft}{\kern0pt}WEST{\isacharunderscore}{\kern0pt}reg{\isacharunderscore}{\kern0pt}aux\ {\isasymphi}\ n{\isacharparenright}{\kern0pt}\ {\isacharparenleft}{\kern0pt}WEST{\isacharunderscore}{\kern0pt}reg{\isacharunderscore}{\kern0pt}aux\ {\isasympsi}\ n{\isacharparenright}{\kern0pt}\ n{\isachardoublequoteclose}\isanewline
{\isacharbar}{\kern0pt}\ {\isachardoublequoteopen}WEST{\isacharunderscore}{\kern0pt}reg{\isacharunderscore}{\kern0pt}aux\ {\isacharparenleft}{\kern0pt}And{\isacharunderscore}{\kern0pt}mltl\ {\isasymphi}\ {\isasympsi}{\isacharparenright}{\kern0pt}\ n\ {\isacharequal}{\kern0pt}\ {\isacharparenleft}{\kern0pt}WEST{\isacharunderscore}{\kern0pt}and{\isacharunderscore}{\kern0pt}simp\ \isanewline
\ \ \ {\isacharparenleft}{\kern0pt}WEST{\isacharunderscore}{\kern0pt}reg{\isacharunderscore}{\kern0pt}aux\ {\isasymphi}\ n{\isacharparenright}{\kern0pt}\ {\isacharparenleft}{\kern0pt}WEST{\isacharunderscore}{\kern0pt}reg{\isacharunderscore}{\kern0pt}aux\ {\isasympsi}\ n{\isacharparenright}{\kern0pt}\ n{\isacharparenright}{\kern0pt}{\isachardoublequoteclose}\isanewline
{\isacharbar}{\kern0pt}\ {\isachardoublequoteopen}WEST{\isacharunderscore}{\kern0pt}reg{\isacharunderscore}{\kern0pt}aux\ {\isacharparenleft}{\kern0pt}Future{\isacharunderscore}{\kern0pt}mltl\ {\isasymphi}\ a\ b{\isacharparenright}{\kern0pt}\ n\ {\isacharequal}{\kern0pt}\ \isanewline
\ \ \ WEST{\isacharunderscore}{\kern0pt}future\ {\isacharparenleft}{\kern0pt}WEST{\isacharunderscore}{\kern0pt}reg{\isacharunderscore}{\kern0pt}aux\ {\isasymphi}\ n{\isacharparenright}{\kern0pt}\ a\ b\ n{\isachardoublequoteclose}\isanewline
{\isacharbar}{\kern0pt}\ {\isachardoublequoteopen}WEST{\isacharunderscore}{\kern0pt}reg{\isacharunderscore}{\kern0pt}aux\ {\isacharparenleft}{\kern0pt}Global{\isacharunderscore}{\kern0pt}mltl\ {\isasymphi}\ a\ b{\isacharparenright}{\kern0pt}\ n\ {\isacharequal}{\kern0pt}\ \isanewline
\ \ \ WEST{\isacharunderscore}{\kern0pt}global\ {\isacharparenleft}{\kern0pt}WEST{\isacharunderscore}{\kern0pt}reg{\isacharunderscore}{\kern0pt}aux\ {\isasymphi}\ n{\isacharparenright}{\kern0pt}\ a\ b\ n{\isachardoublequoteclose}\isanewline
{\isacharbar}{\kern0pt}\ {\isachardoublequoteopen}WEST{\isacharunderscore}{\kern0pt}reg{\isacharunderscore}{\kern0pt}aux\ {\isacharparenleft}{\kern0pt}Until{\isacharunderscore}{\kern0pt}mltl\ {\isasymphi}\ {\isasympsi}\ a\ b{\isacharparenright}{\kern0pt}\ n\ {\isacharequal}{\kern0pt}\ {\isacharparenleft}{\kern0pt}WEST{\isacharunderscore}{\kern0pt}until\ \isanewline
{\isacharparenleft}{\kern0pt}WEST{\isacharunderscore}{\kern0pt}reg{\isacharunderscore}{\kern0pt}aux\ {\isasymphi}\ n{\isacharparenright}{\kern0pt}\ {\isacharparenleft}{\kern0pt}WEST{\isacharunderscore}{\kern0pt}reg{\isacharunderscore}{\kern0pt}aux\ {\isasympsi}\ n{\isacharparenright}{\kern0pt}\ a\ b\ n{\isacharparenright}{\kern0pt}{\isachardoublequoteclose}\isanewline
{\isacharbar}{\kern0pt}\ {\isachardoublequoteopen}WEST{\isacharunderscore}{\kern0pt}reg{\isacharunderscore}{\kern0pt}aux\ {\isacharparenleft}{\kern0pt}Release{\isacharunderscore}{\kern0pt}mltl\ {\isasymphi}\ {\isasympsi}\ a\ b{\isacharparenright}{\kern0pt}\ n\ {\isacharequal}{\kern0pt}\ WEST{\isacharunderscore}{\kern0pt}release\ \isanewline
{\isacharparenleft}{\kern0pt}WEST{\isacharunderscore}{\kern0pt}reg{\isacharunderscore}{\kern0pt}aux\ {\isasymphi}\ n{\isacharparenright}{\kern0pt}\ {\isacharparenleft}{\kern0pt}WEST{\isacharunderscore}{\kern0pt}reg{\isacharunderscore}{\kern0pt}aux\ {\isasympsi}\ n{\isacharparenright}{\kern0pt}\ a\ b\ n{\isachardoublequoteclose}\isanewline
{\isacharbar}{\kern0pt}\ {\isachardoublequoteopen}WEST{\isacharunderscore}{\kern0pt}reg{\isacharunderscore}{\kern0pt}aux\ {\isacharparenleft}{\kern0pt}Not{\isacharunderscore}{\kern0pt}mltl\ True{\isacharunderscore}{\kern0pt}mltl{\isacharparenright}{\kern0pt}\ n\ {\isacharequal}{\kern0pt}\ \isanewline
\ \ \ WEST{\isacharunderscore}{\kern0pt}reg{\isacharunderscore}{\kern0pt}aux\ False{\isacharunderscore}{\kern0pt}mltl\ n{\isachardoublequoteclose}\isanewline
{\isacharbar}{\kern0pt}\ {\isachardoublequoteopen}WEST{\isacharunderscore}{\kern0pt}reg{\isacharunderscore}{\kern0pt}aux\ {\isacharparenleft}{\kern0pt}Not{\isacharunderscore}{\kern0pt}mltl\ False{\isacharunderscore}{\kern0pt}mltl{\isacharparenright}{\kern0pt}\ n\ {\isacharequal}{\kern0pt}\ \isanewline
\ \ \ WEST{\isacharunderscore}{\kern0pt}reg{\isacharunderscore}{\kern0pt}aux\ True{\isacharunderscore}{\kern0pt}mltl\ n{\isachardoublequoteclose}\isanewline
{\isacharbar}{\kern0pt}\ {\isachardoublequoteopen}WEST{\isacharunderscore}{\kern0pt}reg{\isacharunderscore}{\kern0pt}aux\ {\isacharparenleft}{\kern0pt}Not{\isacharunderscore}{\kern0pt}mltl\ {\isacharparenleft}{\kern0pt}And{\isacharunderscore}{\kern0pt}mltl\ {\isasymphi}\ {\isasympsi}{\isacharparenright}{\kern0pt}{\isacharparenright}{\kern0pt}\ n\ {\isacharequal}{\kern0pt}\ \isanewline
\ \ \ WEST{\isacharunderscore}{\kern0pt}reg{\isacharunderscore}{\kern0pt}aux\ {\isacharparenleft}{\kern0pt}Or{\isacharunderscore}{\kern0pt}mltl\ {\isacharparenleft}{\kern0pt}Not{\isacharunderscore}{\kern0pt}mltl\ {\isasymphi}{\isacharparenright}{\kern0pt}\ {\isacharparenleft}{\kern0pt}Not{\isacharunderscore}{\kern0pt}mltl\ {\isasympsi}{\isacharparenright}{\kern0pt}{\isacharparenright}{\kern0pt}\ n{\isachardoublequoteclose}\isanewline
{\isacharbar}{\kern0pt}\ {\isachardoublequoteopen}WEST{\isacharunderscore}{\kern0pt}reg{\isacharunderscore}{\kern0pt}aux\ {\isacharparenleft}{\kern0pt}Not{\isacharunderscore}{\kern0pt}mltl\ {\isacharparenleft}{\kern0pt}Or{\isacharunderscore}{\kern0pt}mltl\ {\isasymphi}\ {\isasympsi}{\isacharparenright}{\kern0pt}{\isacharparenright}{\kern0pt}\ n\ {\isacharequal}{\kern0pt}\ \isanewline
\ \ \ WEST{\isacharunderscore}{\kern0pt}reg{\isacharunderscore}{\kern0pt}aux\ {\isacharparenleft}{\kern0pt}And{\isacharunderscore}{\kern0pt}mltl\ {\isacharparenleft}{\kern0pt}Not{\isacharunderscore}{\kern0pt}mltl\ {\isasymphi}{\isacharparenright}{\kern0pt}\ {\isacharparenleft}{\kern0pt}Not{\isacharunderscore}{\kern0pt}mltl\ {\isasympsi}{\isacharparenright}{\kern0pt}{\isacharparenright}{\kern0pt}\ n{\isachardoublequoteclose}\isanewline
{\isacharbar}{\kern0pt}\ {\isachardoublequoteopen}WEST{\isacharunderscore}{\kern0pt}reg{\isacharunderscore}{\kern0pt}aux\ {\isacharparenleft}{\kern0pt}Not{\isacharunderscore}{\kern0pt}mltl\ {\isacharparenleft}{\kern0pt}Future{\isacharunderscore}{\kern0pt}mltl\ {\isasymphi}\ a\ b{\isacharparenright}{\kern0pt}{\isacharparenright}{\kern0pt}\ n\ {\isacharequal}{\kern0pt}\ \isanewline
\ \ \ WEST{\isacharunderscore}{\kern0pt}reg{\isacharunderscore}{\kern0pt}aux\ {\isacharparenleft}{\kern0pt}Global{\isacharunderscore}{\kern0pt}mltl\ {\isacharparenleft}{\kern0pt}Not{\isacharunderscore}{\kern0pt}mltl\ {\isasymphi}{\isacharparenright}{\kern0pt}\ a\ b{\isacharparenright}{\kern0pt}\ n{\isachardoublequoteclose}\isanewline
{\isacharbar}{\kern0pt}\ {\isachardoublequoteopen}WEST{\isacharunderscore}{\kern0pt}reg{\isacharunderscore}{\kern0pt}aux\ {\isacharparenleft}{\kern0pt}Not{\isacharunderscore}{\kern0pt}mltl\ {\isacharparenleft}{\kern0pt}Global{\isacharunderscore}{\kern0pt}mltl\ {\isasymphi}\ a\ b{\isacharparenright}{\kern0pt}{\isacharparenright}{\kern0pt}\ n\ {\isacharequal}{\kern0pt}\ \isanewline
\ \ \ WEST{\isacharunderscore}{\kern0pt}reg{\isacharunderscore}{\kern0pt}aux\ {\isacharparenleft}{\kern0pt}Future{\isacharunderscore}{\kern0pt}mltl\ {\isacharparenleft}{\kern0pt}Not{\isacharunderscore}{\kern0pt}mltl\ {\isasymphi}{\isacharparenright}{\kern0pt}\ a\ b{\isacharparenright}{\kern0pt}\ n{\isachardoublequoteclose}\isanewline
{\isacharbar}{\kern0pt}\ {\isachardoublequoteopen}WEST{\isacharunderscore}{\kern0pt}reg{\isacharunderscore}{\kern0pt}aux\ {\isacharparenleft}{\kern0pt}Not{\isacharunderscore}{\kern0pt}mltl\ {\isacharparenleft}{\kern0pt}Until{\isacharunderscore}{\kern0pt}mltl\ {\isasymphi}\ {\isasympsi}\ a\ b{\isacharparenright}{\kern0pt}{\isacharparenright}{\kern0pt}\ n\ {\isacharequal}{\kern0pt}\ \isanewline
WEST{\isacharunderscore}{\kern0pt}reg{\isacharunderscore}{\kern0pt}aux\ {\isacharparenleft}{\kern0pt}Release{\isacharunderscore}{\kern0pt}mltl\ {\isacharparenleft}{\kern0pt}Not{\isacharunderscore}{\kern0pt}mltl\ {\isasymphi}{\isacharparenright}{\kern0pt}\ {\isacharparenleft}{\kern0pt}Not{\isacharunderscore}{\kern0pt}mltl\ {\isasympsi}{\isacharparenright}{\kern0pt}\ a\ b{\isacharparenright}{\kern0pt}\ n{\isachardoublequoteclose}\isanewline
{\isacharbar}{\kern0pt}\ {\isachardoublequoteopen}WEST{\isacharunderscore}{\kern0pt}reg{\isacharunderscore}{\kern0pt}aux\ {\isacharparenleft}{\kern0pt}Not{\isacharunderscore}{\kern0pt}mltl\ {\isacharparenleft}{\kern0pt}Release{\isacharunderscore}{\kern0pt}mltl\ {\isasymphi}\ {\isasympsi}\ a\ b{\isacharparenright}{\kern0pt}{\isacharparenright}{\kern0pt}\ n\ {\isacharequal}{\kern0pt}\ \isanewline
WEST{\isacharunderscore}{\kern0pt}reg{\isacharunderscore}{\kern0pt}aux\ {\isacharparenleft}{\kern0pt}Until{\isacharunderscore}{\kern0pt}mltl\ {\isacharparenleft}{\kern0pt}Not{\isacharunderscore}{\kern0pt}mltl\ {\isasymphi}{\isacharparenright}{\kern0pt}\ {\isacharparenleft}{\kern0pt}Not{\isacharunderscore}{\kern0pt}mltl\ {\isasympsi}{\isacharparenright}{\kern0pt}\ a\ b{\isacharparenright}{\kern0pt}\ n{\isachardoublequoteclose}\isanewline
{\isacharbar}{\kern0pt}\ {\isachardoublequoteopen}WEST{\isacharunderscore}{\kern0pt}reg{\isacharunderscore}{\kern0pt}aux\ {\isacharparenleft}{\kern0pt}Not{\isacharunderscore}{\kern0pt}mltl\ {\isacharparenleft}{\kern0pt}Not{\isacharunderscore}{\kern0pt}mltl\ {\isasymphi}{\isacharparenright}{\kern0pt}{\isacharparenright}{\kern0pt}\ n\ {\isacharequal}{\kern0pt}\ \isanewline
\ \ \ WEST{\isacharunderscore}{\kern0pt}reg{\isacharunderscore}{\kern0pt}aux\ {\isasymphi}\ n{\isachardoublequoteclose}%
}
\snip{WESTreg}{1}{2}{%
\isacommand{fun}\isamarkupfalse%
\ WEST{\isacharunderscore}{\kern0pt}reg{\isacharcolon}{\kern0pt}{\isacharcolon}{\kern0pt}\ {\isachardoublequoteopen}nat\ mltl\ {\isasymRightarrow}\ WEST{\isacharunderscore}{\kern0pt}regex{\isachardoublequoteclose}\isanewline
\ \ \isakeyword{where}\ {\isachardoublequoteopen}WEST{\isacharunderscore}{\kern0pt}reg\ {\isasymphi}\ {\isacharequal}{\kern0pt}\ {\isacharparenleft}{\kern0pt}let\ nnf{\isacharunderscore}{\kern0pt}{\isasymphi}\ {\isacharequal}{\kern0pt}\ convert{\isacharunderscore}{\kern0pt}nnf\ {\isasymphi}\ in\ \isanewline
\ \ \ \ \ \ \ \ \ \ \ \ \ \ WEST{\isacharunderscore}{\kern0pt}reg{\isacharunderscore}{\kern0pt}aux\ nnf{\isacharunderscore}{\kern0pt}{\isasymphi}\ {\isacharparenleft}{\kern0pt}WEST{\isacharunderscore}{\kern0pt}num{\isacharunderscore}{\kern0pt}vars\ {\isasymphi}{\isacharparenright}{\kern0pt}{\isacharparenright}{\kern0pt}{\isachardoublequoteclose}%
}
\snip{padWESTreg}{1}{2}{%
\isacommand{fun}\isamarkupfalse%
\ pad{\isacharunderscore}{\kern0pt}WEST{\isacharunderscore}{\kern0pt}reg{\isacharcolon}{\kern0pt}{\isacharcolon}{\kern0pt}\ {\isachardoublequoteopen}nat\ mltl\ {\isasymRightarrow}\ WEST{\isacharunderscore}{\kern0pt}regex{\isachardoublequoteclose}\isanewline
\ \ \isakeyword{where}\ {\isachardoublequoteopen}pad{\isacharunderscore}{\kern0pt}WEST{\isacharunderscore}{\kern0pt}reg\ {\isasymphi}\ {\isacharequal}{\kern0pt}\ {\isacharparenleft}{\kern0pt}let\ unpadded\ {\isacharequal}{\kern0pt}\ WEST{\isacharunderscore}{\kern0pt}reg\ {\isasymphi}\ in\isanewline
\ \ \ \ \ \ \ \ \ \ \ \ \ \ \ \ \ \ \ \ \ \ \ \ \ \ {\isacharparenleft}{\kern0pt}let\ complen\ {\isacharequal}{\kern0pt}\ complen{\isacharunderscore}{\kern0pt}mltl\ {\isasymphi}\ in\isanewline
\ \ \ \ \ \ \ \ \ \ \ \ \ \ \ \ \ \ \ \ \ \ \ \ \ \ {\isacharparenleft}{\kern0pt}let\ num{\isacharunderscore}{\kern0pt}vars\ {\isacharequal}{\kern0pt}\ WEST{\isacharunderscore}{\kern0pt}num{\isacharunderscore}{\kern0pt}vars\ {\isasymphi}\ in\ \isanewline
\ \ {\isacharparenleft}{\kern0pt}map\ {\isacharparenleft}{\kern0pt}{\isasymlambda}\ L{\isachardot}{\kern0pt}\ {\isacharparenleft}{\kern0pt}if\ {\isacharparenleft}{\kern0pt}length\ L\ {\isacharless}{\kern0pt}\ complen{\isacharparenright}{\kern0pt}\ then\ \isanewline
\ \ {\isacharparenleft}{\kern0pt}pad\ L\ num{\isacharunderscore}{\kern0pt}vars\ {\isacharparenleft}{\kern0pt}complen{\isacharminus}{\kern0pt}{\isacharparenleft}{\kern0pt}length\ L{\isacharparenright}{\kern0pt}{\isacharparenright}{\kern0pt}{\isacharparenright}{\kern0pt}\ else\ L{\isacharparenright}{\kern0pt}{\isacharparenright}{\kern0pt}{\isacharparenright}{\kern0pt}\ unpadded{\isacharparenright}{\kern0pt}{\isacharparenright}{\kern0pt}{\isacharparenright}{\kern0pt}{\isachardoublequoteclose}%
}
\snip{simppadWESTreg}{1}{2}{%
\isacommand{fun}\isamarkupfalse%
\ simp{\isacharunderscore}{\kern0pt}pad{\isacharunderscore}{\kern0pt}WEST{\isacharunderscore}{\kern0pt}reg{\isacharcolon}{\kern0pt}{\isacharcolon}{\kern0pt}\ {\isachardoublequoteopen}nat\ mltl\ {\isasymRightarrow}\ WEST{\isacharunderscore}{\kern0pt}regex{\isachardoublequoteclose}\isanewline
\ \ \isakeyword{where}\ {\isachardoublequoteopen}simp{\isacharunderscore}{\kern0pt}pad{\isacharunderscore}{\kern0pt}WEST{\isacharunderscore}{\kern0pt}reg\ {\isasymphi}\ {\isacharequal}{\kern0pt}\ WEST{\isacharunderscore}{\kern0pt}simp\ {\isacharparenleft}{\kern0pt}pad{\isacharunderscore}{\kern0pt}WEST{\isacharunderscore}{\kern0pt}reg\ {\isasymphi}{\isacharparenright}{\kern0pt}\ {\isacharparenleft}{\kern0pt}WEST{\isacharunderscore}{\kern0pt}num{\isacharunderscore}{\kern0pt}vars\ {\isasymphi}{\isacharparenright}{\kern0pt}{\isachardoublequoteclose}%
}
\snip{regtraceofvars}{1}{2}{%
\isacommand{definition}\isamarkupfalse%
\ trace{\isacharunderscore}{\kern0pt}regex{\isacharunderscore}{\kern0pt}of{\isacharunderscore}{\kern0pt}vars{\isacharcolon}{\kern0pt}{\isacharcolon}{\kern0pt}{\isachardoublequoteopen}trace{\isacharunderscore}{\kern0pt}regex\ {\isasymRightarrow}\ nat\ {\isasymRightarrow}\ bool{\isachardoublequoteclose}\isanewline
\isakeyword{where}\ {\isachardoublequoteopen}trace{\isacharunderscore}{\kern0pt}regex{\isacharunderscore}{\kern0pt}of{\isacharunderscore}{\kern0pt}vars\ r\ n\ {\isacharequal}{\kern0pt}{\isacharparenleft}{\kern0pt}{\isasymforall}i{\isacharless}{\kern0pt}length\ r{\isachardot}{\kern0pt}\ length\ {\isacharparenleft}{\kern0pt}r{\isacharbang}{\kern0pt}i{\isacharparenright}{\kern0pt}\ {\isacharequal}{\kern0pt}\ n{\isacharparenright}{\kern0pt}{\isachardoublequoteclose}%
}
\snip{regtraceListofvars}{1}{2}{%
\isacommand{definition}\isamarkupfalse%
\ WEST{\isacharunderscore}{\kern0pt}regex{\isacharunderscore}{\kern0pt}of{\isacharunderscore}{\kern0pt}vars{\isacharcolon}{\kern0pt}{\isacharcolon}{\kern0pt}{\isachardoublequoteopen}WEST{\isacharunderscore}{\kern0pt}regex\ {\isasymRightarrow}\ nat\ {\isasymRightarrow}\ bool{\isachardoublequoteclose}\isanewline
\isakeyword{where}\ {\isachardoublequoteopen}WEST{\isacharunderscore}{\kern0pt}regex{\isacharunderscore}{\kern0pt}of{\isacharunderscore}{\kern0pt}vars\ traceList\ n\ {\isacharequal}{\kern0pt}\ \isanewline
{\isacharparenleft}{\kern0pt}{\isasymforall}k{\isacharless}{\kern0pt}length\ traceList{\isachardot}{\kern0pt}\ trace{\isacharunderscore}{\kern0pt}regex{\isacharunderscore}{\kern0pt}of{\isacharunderscore}{\kern0pt}vars\ {\isacharparenleft}{\kern0pt}traceList{\isacharbang}{\kern0pt}k{\isacharparenright}{\kern0pt}\ n{\isacharparenright}{\kern0pt}{\isachardoublequoteclose}%
}
\snip{matchtimestep}{1}{2}{%
\isacommand{definition}\isamarkupfalse%
\ match{\isacharunderscore}{\kern0pt}timestep{\isacharcolon}{\kern0pt}{\isacharcolon}{\kern0pt}\ {\isachardoublequoteopen}nat\ set\ {\isasymRightarrow}\ state{\isacharunderscore}{\kern0pt}regex\ {\isasymRightarrow}\ bool{\isachardoublequoteclose}\isanewline
\ \ \isakeyword{where}\ {\isachardoublequoteopen}match{\isacharunderscore}{\kern0pt}timestep\ state\ r\ {\isacharequal}{\kern0pt}\ {\isacharparenleft}{\kern0pt}{\isasymforall}i\ {\isacharless}{\kern0pt}\ length\ r{\isachardot}{\kern0pt}\ \isanewline
{\isacharparenleft}{\kern0pt}r\ {\isacharbang}{\kern0pt}\ i\ {\isacharequal}{\kern0pt}\ One\ {\isasymlongrightarrow}\ i\ {\isasymin}\ state{\isacharparenright}{\kern0pt}\ {\isasymand}\ {\isacharparenleft}{\kern0pt}r\ {\isacharbang}{\kern0pt}\ i\ {\isacharequal}{\kern0pt}\ Zero\ {\isasymlongrightarrow}\ i\ {\isasymnotin}\ state{\isacharparenright}{\kern0pt}{\isacharparenright}{\kern0pt}{\isachardoublequoteclose}%
}
\snip{matchregex}{1}{2}{%
\isacommand{definition}\isamarkupfalse%
\ match{\isacharunderscore}{\kern0pt}regex{\isacharcolon}{\kern0pt}{\isacharcolon}{\kern0pt}\ {\isachardoublequoteopen}trace\ {\isasymRightarrow}\ trace{\isacharunderscore}{\kern0pt}regex\ {\isasymRightarrow}\ bool{\isachardoublequoteclose}\isanewline
\isakeyword{where}\ {\isachardoublequoteopen}match{\isacharunderscore}{\kern0pt}regex\ {\isasympi}\ r\ {\isacharequal}{\kern0pt}\ {\isacharparenleft}{\kern0pt}{\isacharparenleft}{\kern0pt}{\isasymforall}\ time{\isacharless}{\kern0pt}length\ r{\isachardot}{\kern0pt}\ \isanewline
{\isacharparenleft}{\kern0pt}match{\isacharunderscore}{\kern0pt}timestep\ {\isacharparenleft}{\kern0pt}{\isasympi}\ {\isacharbang}{\kern0pt}\ time{\isacharparenright}{\kern0pt}\ {\isacharparenleft}{\kern0pt}r\ {\isacharbang}{\kern0pt}\ time{\isacharparenright}{\kern0pt}{\isacharparenright}{\kern0pt}{\isacharparenright}{\kern0pt}{\isasymand}{\isacharparenleft}{\kern0pt}length\ {\isasympi}\ {\isasymge}\ length\ r{\isacharparenright}{\kern0pt}{\isacharparenright}{\kern0pt}{\isachardoublequoteclose}%
}
\snip{match}{1}{2}{%
\isacommand{definition}\isamarkupfalse%
\ match{\isacharcolon}{\kern0pt}{\isacharcolon}{\kern0pt}\ {\isachardoublequoteopen}trace\ {\isasymRightarrow}\ WEST{\isacharunderscore}{\kern0pt}regex\ {\isasymRightarrow}\ bool{\isachardoublequoteclose}\isanewline
\ \ \isakeyword{where}\ {\isachardoublequoteopen}match\ {\isasympi}\ L\ {\isacharequal}{\kern0pt}\ {\isacharparenleft}{\kern0pt}{\isasymexists}i\ {\isacharless}{\kern0pt}\ length\ L{\isachardot}{\kern0pt}\ match{\isacharunderscore}{\kern0pt}regex\ {\isasympi}\ {\isacharparenleft}{\kern0pt}L\ {\isacharbang}{\kern0pt}\ i{\isacharparenright}{\kern0pt}{\isacharparenright}{\kern0pt}{\isachardoublequoteclose}%
}
\snip{regexequiv}{1}{2}{%
\isacommand{definition}\isamarkupfalse%
\ regex{\isacharunderscore}{\kern0pt}equiv{\isacharcolon}{\kern0pt}{\isacharcolon}{\kern0pt}\ {\isachardoublequoteopen}WEST{\isacharunderscore}{\kern0pt}regex\ {\isasymRightarrow}\ WEST{\isacharunderscore}{\kern0pt}regex\ {\isasymRightarrow}\ bool{\isachardoublequoteclose}\isanewline
\ \ \isakeyword{where}\ {\isachardoublequoteopen}regex{\isacharunderscore}{\kern0pt}equiv\ rl{\isadigit{1}}\ rl{\isadigit{2}}\ {\isacharequal}{\kern0pt}\ \isanewline
\ \ {\isacharparenleft}{\kern0pt}{\isacharparenleft}{\kern0pt}{\isasymforall}\ {\isasympi}{\isacharcolon}{\kern0pt}{\isacharcolon}{\kern0pt}trace{\isachardot}{\kern0pt}\ {\isacharparenleft}{\kern0pt}match\ {\isasympi}\ rl{\isadigit{1}}{\isacharparenright}{\kern0pt}\ {\isasymlongleftrightarrow}\ {\isacharparenleft}{\kern0pt}match\ {\isasympi}\ rl{\isadigit{2}}{\isacharparenright}{\kern0pt}{\isacharparenright}{\kern0pt}{\isacharparenright}{\kern0pt}{\isachardoublequoteclose}%
}
\snip{WESTandbitwise}{1}{2}{%
\isacommand{fun}\isamarkupfalse%
\ WEST{\isacharunderscore}{\kern0pt}and{\isacharunderscore}{\kern0pt}bitwise{\isacharcolon}{\kern0pt}{\isacharcolon}{\kern0pt}\ {\isachardoublequoteopen}WEST{\isacharunderscore}{\kern0pt}bit{\isasymRightarrow}WEST{\isacharunderscore}{\kern0pt}bit{\isasymRightarrow}WEST{\isacharunderscore}{\kern0pt}bit option{\isachardoublequoteclose} \isakeyword{where}\isanewline
{\isachardoublequoteopen}WEST{\isacharunderscore}{\kern0pt}and{\isacharunderscore}{\kern0pt}bitwise\ b\ One\ {\isacharequal}{\kern0pt}\ {\isacharparenleft}{\kern0pt}if\ b{\isacharequal}{\kern0pt}Zero\ then\ None\ else\ Some\ One{\isacharparenright}{\kern0pt}{\isachardoublequoteclose}\isanewline
{\isacharbar}{\kern0pt}\ {\isachardoublequoteopen}WEST{\isacharunderscore}{\kern0pt}and{\isacharunderscore}{\kern0pt}bitwise\ b\ Zero\ {\isacharequal}{\kern0pt}\ {\isacharparenleft}{\kern0pt}if\ b{\isacharequal}{\kern0pt}One\ then\ None\ else\ Some\ Zero{\isacharparenright}{\kern0pt}{\isachardoublequoteclose}\ \isanewline
{\isacharbar}{\kern0pt}\ {\isachardoublequoteopen}WEST{\isacharunderscore}{\kern0pt}and{\isacharunderscore}{\kern0pt}bitwise\ b\ S\ {\isacharequal}{\kern0pt}\ Some\ b{\isachardoublequoteclose}%
}
\snip{mltlsyntax}{1}{2}{%
\isacommand{datatype}\isamarkupfalse%
\ {\isacharparenleft}{\kern0pt}atoms{\isacharunderscore}{\kern0pt}mltl{\isacharcolon}{\kern0pt}\ {\isacharprime}{\kern0pt}a{\isacharparenright}{\kern0pt}\ mltl\ {\isacharequal}{\kern0pt}\isanewline
\ \ \ \ True{\isacharunderscore}{\kern0pt}mltl\ {\isacharbar}{\kern0pt}\ False{\isacharunderscore}{\kern0pt}mltl\ {\isacharbar}{\kern0pt}\ Prop{\isacharunderscore}{\kern0pt}mltl\ {\isacharprime}{\kern0pt}a\ \isanewline
\ \ {\isacharbar}{\kern0pt}\ Not{\isacharunderscore}{\kern0pt}mltl\ {\isachardoublequoteopen}{\isacharprime}{\kern0pt}a\ mltl{\isachardoublequoteclose}\ {\isacharbar}{\kern0pt}\ And{\isacharunderscore}{\kern0pt}mltl\ {\isachardoublequoteopen}{\isacharprime}{\kern0pt}a\ mltl{\isachardoublequoteclose}\ {\isachardoublequoteopen}{\isacharprime}{\kern0pt}a\ mltl{\isachardoublequoteclose}\ \isanewline
\ \ {\isacharbar}{\kern0pt}\ Or{\isacharunderscore}{\kern0pt}mltl\ {\isachardoublequoteopen}{\isacharprime}{\kern0pt}a\ mltl{\isachardoublequoteclose}\ {\isachardoublequoteopen}{\isacharprime}{\kern0pt}a\ mltl{\isachardoublequoteclose}\ \ \ \ \ \ \ \ \ \ \ \ \isanewline
\ \ {\isacharbar}{\kern0pt}\ Future{\isacharunderscore}{\kern0pt}mltl\ {\isachardoublequoteopen}{\isacharprime}{\kern0pt}a\ mltl{\isachardoublequoteclose}\ {\isachardoublequoteopen}nat{\isachardoublequoteclose}\ {\isachardoublequoteopen}nat{\isachardoublequoteclose}\ \isanewline
\ \ {\isacharbar}{\kern0pt}\ Global{\isacharunderscore}{\kern0pt}mltl\ {\isachardoublequoteopen}{\isacharprime}{\kern0pt}a\ mltl{\isachardoublequoteclose}\ {\isachardoublequoteopen}nat{\isachardoublequoteclose}\ {\isachardoublequoteopen}nat{\isachardoublequoteclose}\ \ \ \ \ \ \ \isanewline
\ \ {\isacharbar}{\kern0pt}\ Until{\isacharunderscore}{\kern0pt}mltl\ {\isachardoublequoteopen}{\isacharprime}{\kern0pt}a\ mltl{\isachardoublequoteclose}\ {\isachardoublequoteopen}{\isacharprime}{\kern0pt}a\ mltl{\isachardoublequoteclose}\ {\isachardoublequoteopen}nat{\isachardoublequoteclose}\ {\isachardoublequoteopen}nat{\isachardoublequoteclose}\ \ \ \ \ \ \ \isanewline
\ \ {\isacharbar}{\kern0pt}\ Release{\isacharunderscore}{\kern0pt}mltl\ {\isachardoublequoteopen}{\isacharprime}{\kern0pt}a\ mltl{\isachardoublequoteclose}\ {\isachardoublequoteopen}{\isacharprime}{\kern0pt}a\ mltl{\isachardoublequoteclose}\ {\isachardoublequoteopen}nat{\isachardoublequoteclose}\ {\isachardoublequoteopen}nat{\isachardoublequoteclose}%
}
\snip{WESTdatatype}{1}{2}{%
\isacommand{datatype}\isamarkupfalse%
\ WEST{\isacharunderscore}{\kern0pt}bit\ {\isacharequal}{\kern0pt}\ Zero\ {\isacharbar}{\kern0pt}\ One\ {\isacharbar}{\kern0pt}\ S%
}
\snip{traceexample}{1}{2}{%
\isacommand{value}\isamarkupfalse%
\ {\isachardoublequoteopen}{\isacharbrackleft}{\kern0pt}{\isacharbraceleft}{\kern0pt}{\isadigit{0}}{\isacharcomma}{\kern0pt}\ {\isadigit{1}}{\isacharbraceright}{\kern0pt}{\isacharcomma}{\kern0pt}\ {\isacharbraceleft}{\kern0pt}{\isadigit{0}}{\isacharbraceright}{\kern0pt}{\isacharbrackright}{\kern0pt}{\isacharcolon}{\kern0pt}{\isacharcolon}{\kern0pt}trace{\isachardoublequoteclose}%
}

\begin{document}
\title{Formally Verifying a Transformation from MLTL Formulas to Regular Expressions}
\titlerunning{Verifying MLTL to Regular Expressions}
%

\author{Zili Wang\inst{1}\orcidID{0000-0003-1730-6180} \and
Katherine Kosaian\inst{2}\orcidID{0000-0002-9336-6006} \and
Kristin Yvonne Rozier\inst{1}\orcidID{0000-0002-6718-2828}}
\authorrunning{Wang et al.}

\institute{Iowa State University, Ames, IA, USA\and
University of Iowa, Iowa City, IA, USA\\
\email{\{ziliw1,kyrozier\}@iastate.edu},
\email{katherine-kosaian@uiowa.edu}}

\maketitle              
\begin{abstract}
Mission-time Linear Temporal Logic (MLTL), a widely used subset of popular specification logics like STL and MTL, is often used to model and verify real world systems in safety-critical contexts. 
As the results of formal verification are only as trustworthy as their input specifications, the WEST tool was created to facilitate writing MLTL specifications. Accordingly, it is vital to demonstrate that WEST itself works correctly. 
To that end, we verify the WEST algorithm, which converts MLTL formulas to (logically equivalent) regular expressions, in the theorem prover Isabelle/HOL.
Our top-level result establishes the correctness of the regular expression transformation; we then generate a code export from our verified development and use this to experimentally validate the existing WEST tool.
To facilitate this, we develop some verified support for checking the equivalence of two regular expressions.

\end{abstract}
\section{Introduction}

As formal methods tools become increasingly integrated into system development life cycles, it is necessary to offer stronger demonstrations of their correct implementation than piecemeal code analysis and experimental validation.
After all, these are the tools justifying and verifying, e.g., the certification of systems; these tools must obey a higher standard for correctness.
This starts with their input languages and specification validation. 

Many formal methods tools, such as model checkers and runtime verification engines, reason over behavior specifications in LTL or related linear-time logics that extend LTL, e.g., to add intervals on the temporal operators like Signal Temporal Logic (STL) \cite{MN04}, Metric Temporal Logic (MTL) \cite{AH90}, and Metric Interval Temporal Logic (MITL) \cite{DBLP:conf/podc/AlurFH91}. Mission-time Linear Temporal Logic (MLTL) \cite{RRS14,LVR19} represents a commonly used subset of these timed logics, and has a conversion to LTL \cite{LVR19}. Several tools use MLTL as a core specification language; these include the Formal Requirements Elicitation Tool (FRET)  \cite{GMRPSS20,NASA-FRET,Mav22}, the Realizable Responsive Unobtrusive Unit (R2U2) \cite{RRS14,RS17,JJKRZ23}, and the Ogda runtime monitoring tool \cite{PMPGG22,Per23,ogma}. 
Popular symbolic model checker \textsc{nuXmv} \cite{DBLP:conf/cav/CavadaCDGMMMRT14} supports a subset of MLTL \cite{nuXmv-v1.1.0} by allowing bounds on the Globally and Finally operators (but not on Until or Release).
The WEST tool \cite{DBLP:conf/ifm/ElwingGSTWR23,WEST2024tool} transforms MLTL formulas into logically equivalent (and easier to analyze) regular expressions and facilitates the validation of MLTL specifications with an interactive GUI.  
Since WEST validates specifications, which are the fundamental basis for formal verification, it is especially critical to rigorously establish its correctness.

The research community has long recognized that specification is the biggest bottleneck in formal methods \cite{Roz16}; to that end LTL is formalized in Coq \cite{DBLP:journals/logcom/Coupet-Grimal03}, in PVS \cite{DBLP:conf/birthday/PnueliA03}, and in Isabelle/HOL \cite{LTL-AFP}, 
along with many algorithms for its use in formal verification \cite{LTL_Normal_Form-AFP,LTL_Master_Theorem-AFP,LTL_to_DRA-AFP,LTL_to_GBA-AFP,CAVA_LTL_Modelchecker-AFP}. 
Libraries for related linear-time logics were inspired by, or directly built upon those for LTL, including formalizations of MTL in Coq \cite{OnlineMTL_Coq} and PVS \cite{DBLP:conf/cpp/ConradTGPD22,fret-proof-framework}; 
a PVS formalization of MITL \cite{DBLP:conf/vmcai/Roohi018}; 
and Isabelle formalizations of the 3-valued variant LTL3 \cite{LTL3_Semantics-AFP} and MLTL \cite{KWSR24}.
Further, the importance of ensuring correctness of formal methods tools naturally prompts using these formalizations to generate tools.
For instance, an Isabelle/HOL formalization of the VeriMon tool for monitoring metric first-order temporal logic (MFOTL) generates (via code export) VeriMon's codebase \cite{DBLP:conf/ictac/BasinDHHMKKMRST22}. 
An Isabelle/HOL formalization of a metric dynamic logic (MDL) runtime monitoring algorithm also generated the Vydra tool \cite{raszyk2020multi}.\footnote{Vydra also reasons with regular expressions in the input language, rather than using regular expression to represent the input, as WEST does.}
In Coq, a formalization of monitoring past-time MTL generates an OCaml monitoring engine \cite{MTL_monitor_coq}.

We enrich this space by formalizing the WEST algorithm for specification validation. 
Building on an existing MLTL library in Isabelle/HOL \cite{KWSR24,Mission_Time_LTL-AFP}, we formally prove that the WEST algorithm generates regular expressions that are logically equivalent to the input MLTL formulas, filling in details omitted from the original tool's correctness proofs. 
From our formalized algorithms, we generate a new implementation of WEST to validate the (unverified) implementations of WEST: the proof-of-concept original \cite{DBLP:conf/ifm/ElwingGSTWR23} and a highly optimized refactoring \cite{WEST2024tool}.
As WEST validates other MLTL tools, most notably the runtime verification engine R2U2 \cite{RRS14}, our work helps to foster trust in a safety-critical space.
Our experiments also show that our Isabelle-generated code is (in aggregate) close in performance to the optimized, unverified version of WEST. 

Section \ref{sec:prelim} recaps the existing Isabelle/HOL MLTL library \cite{KWSR24}, introduces the trace regular expressions fundamental to the WEST algorithm, and sets up the definitions underlying our formalization. 
Section \ref{sec:formal} presents our formalization of the WEST algorithm. 
Section \ref{sec:insights} gathers our formalization insights to inform future efforts that build on our contributions. 
Section \ref{sec:experiments} experimentally evaluates the new version of WEST generated via Isablle's code export utility in comparison with two previous, hand-coded versions \cite{DBLP:conf/ifm/ElwingGSTWR23,WEST2024tool}, while Section \ref{sec:conclusion} concludes with a discussion.
Our formalization (totaling $\approx$ 7400 lines of code) is available on the Archive of Formal Proofs (AFP) \cite{Mission_Time_LTL_to_Regular_Expression-AFP}.

\section{MLTL and Regular Expressions}
\label{sec:prelim}

In this section, we present the syntax and semantics of MLTL and explain our formalization of the WEST regular expressions used by the WEST tool, highlighting some key datatypes; when appropriate, we intersperse mathematical definitions with Isabelle/HOL code.
We also introduce some useful functions that are important in the correctness proofs later on.

Other works formalize regular expressions in different contexts.
An algorithm for matching extended regular expressions via symbolic derivatives was formalized in Lean \cite{Lean_Extended_regex}, and the Myhill-Nerode theorem was restated in Isabelle/HOL using regular expressions (instead of automata, which is more common) \cite{MyhillNerodeFormalization}. 
There has also been work formalizing decision procedures to check equivalence of regular expressions formalized in Coq \cite{DBLP:conf/cpp/CoquandS11} and Isabelle/HOL \cite{DBLP:journals/jar/KraussN12}.
The latter is particularly relevant; we are interested in potentially incorporating it in future work to improve our (currently naive) regular expression checking procedure.

\subsection{Syntax and Semantics of MLTL}
Let $\texttt{AP}$ be a finite set of atomic propositions. 
Let $\texttt{p} \in \texttt{AP}$ be an atomic proposition, and $a, b \in \mathbb{N}$ be natural numbers such that $a \leq b$; MLTL formulas $\phi$, $\psi$, and $\xi$ are defined recursively as follows; the temporal operators $\texttt{F}, \texttt{G}, \texttt{U}, \texttt{R}$ denote ``Future'', ``Globally'', ``Until'', and ``Release'', respectively.
$$ 
\xi := 
\texttt{True} \ | 
\ \texttt{False} \ | 
\ \texttt{p} \ | 
\ \neg \phi \ | 
\ \phi \land \psi \ | 
\ \phi \lor \psi \ | 
\ \texttt{F}_{[a,b]} \phi \ | 
\ \texttt{G}_{[a,b]} \phi \ | 
\ \phi \texttt{U}_{[a,b]} \psi \ | 
\ \phi \texttt{R}_{[a,b]} \psi.
$$

A \textbf{trace} $\pi$ is a finite sequence $\pi = \pi[0], \pi[1], \ldots$ of sets of atomic propositions, where $\pi[i] \subseteq \texttt{AP}$ for all $i$.
We refer to the $i$-th element of a trace $\pi$ as the $i$-th state of the trace, and intuitively interpret $\pi[i]$ as the set of propositions that are true at time $i$.
We denote the length of a trace $\pi$ by $|\pi|$, and the suffix of a trace $\pi$ starting at time $i$ by $\pi_i$; that is, $\pi_i = \pi[i], \pi[i+1], \ldots$ and $\pi_0 = \pi$.
The existing MLTL library in Isabelle/HOL \cite{Mission_Time_LTL-AFP} encodes a trace as a list of sets of natural numbers; each set represents the atomic propositions that are true at each timestep.
For example, the trace $\pi = \{p_0, p_1\}, \{p_0\}$ is encoded in Isabelle as the \isa{{\isacharbrackleft}{\kern0pt}{\isacharbraceleft}{\kern0pt}{\isadigit{0}}{\isacharcomma}{\kern0pt}\ {\isadigit{1}}{\isacharbraceright}{\kern0pt}{\isacharcomma}{\kern0pt}\ {\isacharbraceleft}{\kern0pt}{\isadigit{0}}{\isacharbraceright}{\kern0pt}{\isacharbrackright}}, which has type \isa{nat set list}.

A trace $\pi$ satisfies an MLTL formula $\phi$, denoted $\pi \models \phi$, as follows \cite{RRS14,LVR19}, where $\psi$ is another MLTL formula:\\
\begin{minipage}{0.5\textwidth}
\begin{flalign*}
    &\pi \models p \text{ iff } p \in \pi[0]&\\
    &\pi \models \phi \land \psi \text{ iff } \pi \models \phi \text{ and } \pi \models \psi&
\end{flalign*}
\end{minipage}
\begin{minipage}{0.5\textwidth}
\begin{flalign*}
    &\pi \models \neg \phi \text{ iff } \pi \not\models \phi&\\
    &\pi \models \phi \lor \psi \text{ iff } \pi \models \phi \text{ or } \pi \models \psi&
\end{flalign*}
\end{minipage}
    {\setlength{\abovedisplayskip}{0pt}
\begin{flalign*}
    &\pi \models \texttt{F}_{[a,b]}\phi \text{ iff } |\pi| > a \text{ and } \exists i \in [a,b].\ \pi_i \models \phi&\\
    &\pi \models \texttt{G}_{[a,b]}\phi \text{ iff } |\pi| \leq a \text{ or } \forall i \in [a,b].\ \pi_i \models \phi&\\
    &\pi \models \phi \ \texttt{U}_{[a,b]} \psi \text{ iff } |\pi| > a \text{ and } \exists i \in [a,b].\ (\pi_i \models \psi \text{ and }\forall j \in [a, i-1].\ \pi_j \models \phi) &\\
    &\pi \models \phi \ \texttt{R}_{[a,b]} \psi \text{ iff } |\pi| \leq a \text{ or } (\forall i \in [a,b]. \ \pi_i \models \psi) \text{ or }\exists j \in [a,b-1].\ (\pi_j \models \phi \text{ and }&\\
    &\quad \quad \quad \quad \quad \quad \quad \forall k \in [a, j] \ \pi_k \models \psi)&
\end{flalign*}}

\subsection{Trace Regular Expressions}\label{sec:TRE}

The WEST algorithm \cite{DBLP:conf/ifm/ElwingGSTWR23} takes an MLTL formula as input and recursively computes a WEST regular expression representing exactly the set of traces that satisfy that formula. 
Intuitively, we can think of this as happening in two steps.
First, we represent traces as \textbf{bit strings}; here, instead of encoding each state in a trace as a set, we encode each state as a bit string of length $n$ (where $n$ is the number of variables in the formula).
Next, we define \textbf{WEST regular expressions (WEST regexes)}, as a compact way to represent a set of traces.

More precisely, we assume that $\texttt{AP} = \{p_0, p_1, \ldots, p_{n-1}\}$ and impose (without loss of generality) an ordering on these atomic propositions; we use this ordering to construct the \textbf{bit string} of a trace $\pi$ of length $m$ as the length $mn$ string of $0$'s and $1$'s such that the value of atomic proposition $p_k$ at timestep $i$ corresponds to the $(ni+k)$-th character of the bit string \cite[Definition 2]{DBLP:conf/ifm/ElwingGSTWR23}.
\begin{wrapfigure}{r}{0.43\textwidth}
        \includegraphics[width=\linewidth]{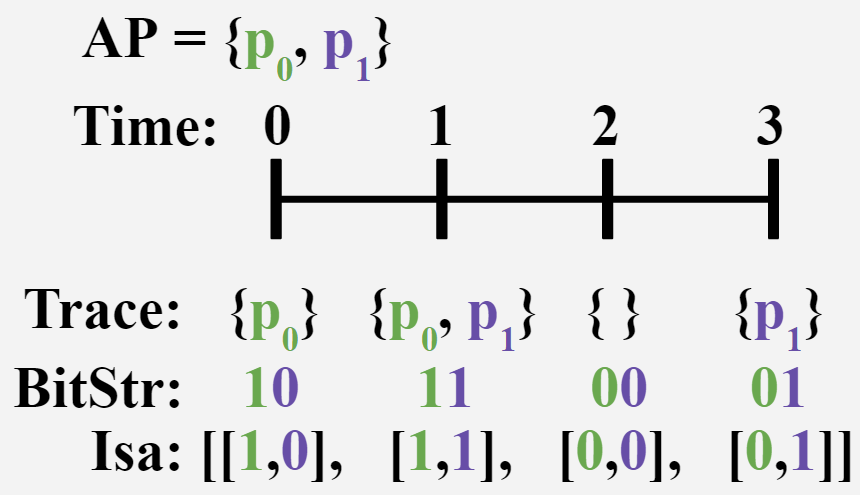}
        \caption{
        For $\texttt{AP} = \{p_0, p_1\}$, the bit string of trace $\{p_0\}, \{p_0, p_1\}, \{\}, \{p_1\}$ is \texttt{10,11,00,01} (following the source material \cite{DBLP:conf/ifm/ElwingGSTWR23}, we use commas to separate timesteps for readability) which is encoded in Isabelle as \isa{[[1,0], [1, 1], [0, 0], [0, 1]]} (type \isa{nat list list}).}
        \label{fig:trace_regex}
            \end{wrapfigure}
We visualize an example in \rref{fig:trace_regex}. 
We encode bit strings in Isabelle as lists of lists.

In Isabelle/HOL, we obtain an ordering on our set of atomic propositions by constraining them to be natural numbers, of type \isa{nat}.
Following WEST's implementation \cite{WEST2024tool}, we choose not to fix $n$ globally (which we could accomplish using a locale \cite{DBLP:conf/types/Ballarin03,DBLP:journals/jar/Ballarin14}) but instead pass the number of variables as an argument to the helper functions in the WEST algorithm (in the top-level function, we compute the right value to pass to the helper functions). 

We then collate these bit string representations in \textbf{trace regular expressions},\footnote{Also called temporal regular expressions \cite[Definition 4]{DBLP:conf/ifm/ElwingGSTWR23}.} or trace regexes for short, which are strings consisting of \texttt{0}, \texttt{1}, and \texttt{S}, where \texttt{S} is a shorthand for the regular expression $\texttt{0} | \texttt{1}$.
For example, fixing the number of atomic propositions to be $n = 3$, the trace regex \texttt{10S} matches only the two bit strings \texttt{101} and \texttt{100} (each representing a trace of length $1$), and the trace regex \texttt{S00,0S0} matches the four bit strings (each representing a length $2$ trace) 
``\texttt{100,010}'', ``\texttt{100,000}'', ``\texttt{000,010}'', and ``\texttt{000,000}''.

In Isabelle/HOL, trace regexes have type \isa{WEST\_bit list list}, where our custom datatype \isa{WEST\_bit} is comprised by \isa{Zero}, \isa{One}, and \isa{S}.
We represent trace regexes with \isa{WEST\_bit list list} and not \isa{WEST\_bit list} because the number of atomic propositions, $n$, is critical for the interpretation of traces from their bit string representations.
We must ensure that each \isa{WEST\_bit list}, referred to as a \textit{state regex}, has length $n$ in the overall list; having a list of lists facilitates this check. 
For this, we define the function \isa{trace\_regex\_of\_vars} which takes as inputs trace regex \isa{r} and the number of atomic propositions \isa{n}, and checks that each state regex in \isa{r} has length \isa{n}.
Here, \isa{!} is Isabelle/HOL syntax for the \isa{i}-th element of \isa{L}.

\begin{mdframed}[backgroundcolor=black!10,linecolor=black!10]
\begin{isa}
    \regtraceofvars
\end{isa}
\end{mdframed}

Then, we build a list of trace regexes as a \isa{WEST\_regex} of type \isa{WEST\_bit list list list}, the final return type of the WEST algorithm.
A WEST regex \isa{L} is well-defined for $n$ atomic propositions if each trace regex \isa{r} in \isa{L} satisfies \isa{trace\_regex\_of\_vars r n}.
We summarize the datatypes of objects in our encoding in \rref[]{tab:type_summary}.
While the nested lists may seem unwieldy at first glance, they ensure modularity in the implementation and, more crucially, in the correctness proofs. 
We turn to an example of this modularity now, as we build up to formalizing the notion of a WEST regex matching a trace.

\begin{table}[]
    \centering
    \resizebox{\columnwidth}{!}{%
    \begin{tabular}{|l|l|l|}
    \hline
    \textbf{Terminology} & \textbf{Description} & \textbf{Isabelle Type} \\ \hline
    WEST bit & Custom Isabelle datatype & \isa{WEST\_bit} \\ \hline
    state regex & List of WEST bits that encodes states as bit strings & \isa{WEST\_bit list} \\ \hline
    trace regex & \begin{tabular}[c]{@{}l@{}}List of WEST states that represents \\ sets of traces compactly as regular expressions\end{tabular} & \isa{WEST\_bit list list} \\ \hline
    WEST regex & \begin{tabular}[c]{@{}l@{}}List of WEST traces that represents the union of \\ all sets of traces represented by the WEST traces\end{tabular} & \isa{WEST\_bit list list list} \\ \hline
    \end{tabular}%
    }
    \caption{Summary of the datatypes of each object in our encoding.}
    \label{tab:type_summary}
\end{table}

\subsection{Useful Definitions}
The notion of matching is foundational to the WEST algorithm because it is crucial for connecting the semantics of MLTL formulas to the semantics of WEST regexes.
We define that a state regex $r$ \textbf{matches} a state if $r$ equals the bit string representation of the state or if $r$ generalizes the bit string by replacing some characters in the bit string with \texttt{S}'s.
This notion lifts to traces: a trace regex $r$ matches a trace $\pi$ iff $r$ matches the bit string representation of $\pi$.
Furthermore, we may lift this to WEST regexes. For trace regexes $r_1, r_2, ..., r_k$, we can combine them by alternations as $r_1 | r_2 | ... | r_k$; we abbreviate this as the WEST regex $L = [r_1, r_2, ..., r_k]$, and define that $L$ matches a trace $\pi$ iff some $r_i$ matches $\pi$.

We contribute a formal mathematical definition of the notion of matching, which previous work \cite{DBLP:conf/ifm/ElwingGSTWR23} supplied only an intuition for.
We do this in three steps.
First, we define matching a state regex (of type \isa{WEST\_bit list}) to a state in a trace (of type \isa{nat\ set}) in the definition \isa{match\_timestep}:
\begin{mdframed}[backgroundcolor=black!10,linecolor=black!10]
\begin{isa}
    \matchtimestep
\end{isa}
\end{mdframed}
This definition checks that for all \isa{i}, \isa{r!i} equaling \isa{One} implies the $i$-th atomic proposition $p_i$ holds at the input state (i.e., $p_i \in \isa{state}$), and \isa{r!i} equaling \isa{Zero} implies $p_i$ does not hold at this state.
If \isa{r!i} is \isa{S}, then $p_i$ can be either true or false at this state.
For example, the state regex $[0, 1, S]$ matches $\{1\}$ and $\{1, 2\}$.

Next we define matching a trace regex (of type \isa{WEST\_bit list list}) to a trace (of type \isa{nat\ set\ list}) in \isa{match\_regex}:
\begin{mdframed}[backgroundcolor=black!10,linecolor=black!10]
\begin{isa}
    \matchregex
\end{isa}
\end{mdframed}
This definition takes as input a trace \isasympi\ and a trace regex \isa{r}, and checks that \isa{match\_timestep} holds for all regex states in \isa{trace} (i.e., for all \isa{r\ !\ time}) on the corresponding state in the trace (\isa{\isasympi\ !\ time}).
It also checks that the length of \isasympi\ is at least the length of \isa{r} (a well-definedness condition, as we need to access \isa{\isasympi\ !\ time} for all time up to the length of \isa{r}).

Finally, we define matching a WEST regex (of type \isa{WEST\_bit list list list}) to a trace (of type \isa{nat\ set\ list}) in the definition \isa{match}:

\begin{mdframed}[backgroundcolor=black!10,linecolor=black!10]
\begin{isa}
    \match
\end{isa}
\end{mdframed}
This definition checks that \isa{match\_regex} holds for some trace regex \isa{L ! i} in \isa{L} and the trace \isasympi.
We may intuitively view WEST regexes as compactly representing the behavior of a set of traces; then, the WEST algorithm transforms a given MLTL formula into a WEST regex that captures the set of satisfying traces. 

Another important function, \isa{WEST\_num\_vars}, counts the number of atomic propositions in a given MLTL formula by recursively computing the maximum number of atomic propositions in all subformulas.
For example, \isa{WEST\_num\_vars} of an atomic proposition \isa{p} is \isa{p+1} (as atomic propositions are indexed from $0$), and \isa{WEST\_num\_vars} of \isa{And\_mltl}\ \isasymphi\ \isasympsi\ is the maximum of \isa{WEST\_num\_vars} \isasymphi\ and \isa{WEST\_num\_vars} \isasympsi.
This function is used frequently in our correctness results.

\begin{wrapfigure}{r}{0.44\textwidth}
    \centering
    \includegraphics[width=\linewidth]{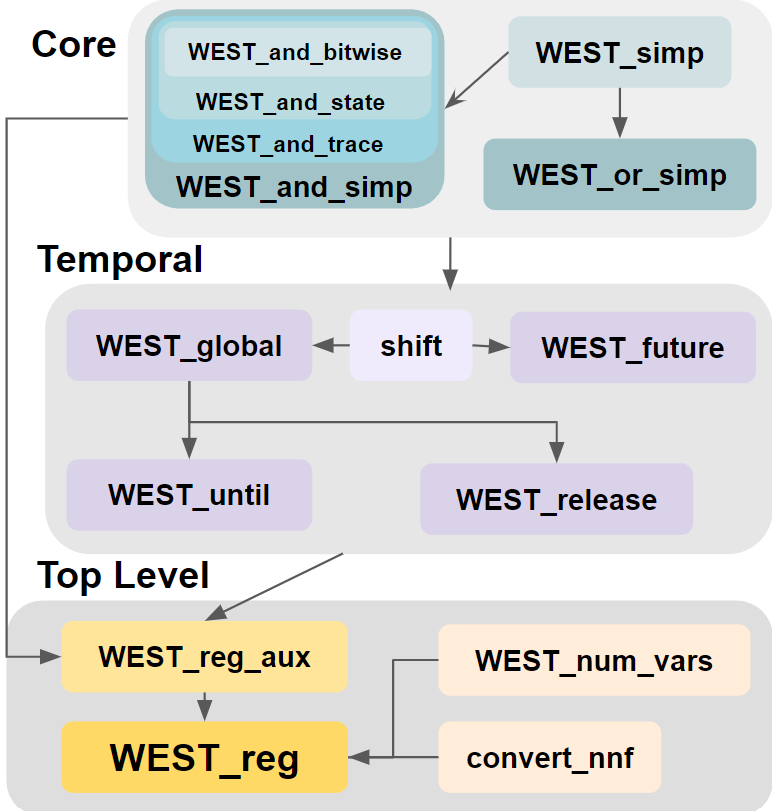}
    \caption{High-level overview of key components in our formalization of the WEST algorithm.
    }
    \label{fig:west_alg}
\end{wrapfigure}

\section{Formalizing the WEST Algorithm}
\label{sec:formal}
Intuitively, the WEST algorithm recursively computes a list of trace regexes for the subformulas of an MLTL formula, and then combines these lists using the \texttt{WEST\_and} and \texttt{WEST\_or} operations for taking intersections and unions of sets of traces. 
The finite semantics of MLTL formulas ensures that all existential and universal quantifiers can be translated to a finite number of \texttt{WEST\_and} and \texttt{WEST\_or} operations on trace regexes; thus the WEST algorithm directly defines the temporal operators in terms of \texttt{WEST\_and} and \texttt{WEST\_or}.
For these temporal operators, we also need a \textit{shifting} operation, \isa{shift}, which the source material \cite{DBLP:conf/ifm/ElwingGSTWR23} implicitly uses but does not explicitly define.
Intuitively, \isa{shift} ensures that we are analyzing the locations in the trace specified by the temporal operators; we will see this in an example in \rref{sec:temporal}.
\rref{fig:west_alg} visualizes the overall structure of the WEST algorithm.

We first discuss our formalization of the core operators \isa{WEST\_and} and \isa{WEST\_or} along with our formalization of an important simplification step in \rref{sec:core}.
Then, we present how the temporal operators 
are built on top of these core operators in \rref{sec:temporal}, using the \isa{shift} operation.
Finally, we discuss the top-level WEST algorithm \isa{WEST\_reg} and our overall correctness result in \rref{sec:toplevel}.

\subsection{The Core Operations of WEST}\label{sec:core}

The \texttt{WEST\_or} operation simply combines two WEST regexes (i.e., lists of trace regexes) into one WEST regex.
We implement this in Isabelle/HOL using the built-in \isa{@} operator for list concatenation.
The top-level correctness theorem shows that for two WEST regexes \isa{L1} and \isa{L2}, \isa{L1} matches a trace \isasympi\ or \isa{L2} matches \isasympi\ iff \isa{L1@L2} matches \isasympi.
We formally state this as the \isa{WEST\_or\_correct} lemma.

\begin{mdframed}[backgroundcolor=black!10,linecolor=black!10]
\begin{isa}
    \WESTorcorrect
\end{isa}
\end{mdframed}

\begin{figure}
\centering
\includegraphics[width=\linewidth]{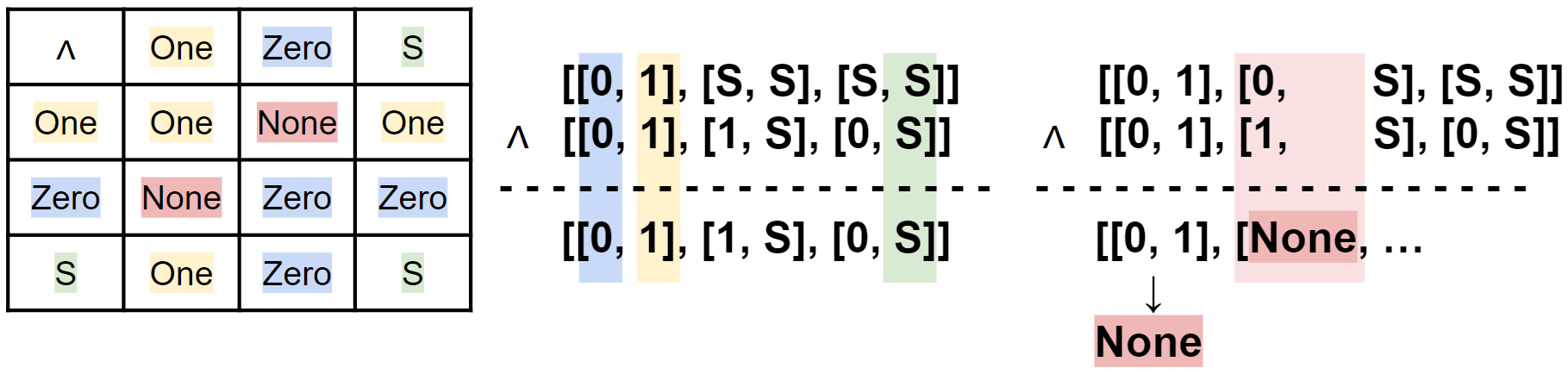}
\caption{Operations table for \texttt{WEST\_and} operation for bits (left), and two examples of \texttt{WEST\_and} between regex states and traces (middle and right).}
\label{fig:and_ex}
\end{figure}

Next, the \texttt{WEST\_and} operation takes as input two lists of trace regexes and computes a list of trace regexes representing the intersection of the sets of traces represented by the input lists.
We visualize the intended semantics of this operation in \rref{fig:and_ex}.
One notable point here is that \isa{WEST\_and\ Zero\ One} is \isa{None}, because it is impossible for a bit in a trace regex to simultaneously equal \isa{Zero} and \isa{One}.
In Isabelle/HOL, we formalize \texttt{WEST\_and} in four steps: first we define an operation between two bits, then between two regex states, then between two trace regexes, and finally between two WEST regexes.

The lowest-level operation between two bits (each of type \isa{WEST\_bit}) is defined in the function \isa{WEST\_and\_bitwise} as follows:

\begin{mdframed}[backgroundcolor=black!10,linecolor=black!10]
\noindent\begin{isa}
    \WESTandbitwise
\end{isa}
\end{mdframed}

This operation reflects the desired semantics visualized in \rref{fig:and_ex} by using option types to return \isa{None} when the set intersection is empty.
For example, \isa{WEST\_and\_bitwise S Zero} is \isa{Some\ Zero}, while \isa{WEST\_and\_bitwise One Zero} is \isa{None}. 

This operation is then lifted to two regex states in \isa{WEST\_and\_state}; here, we apply \isa{WEST\_and\_bitwise} to each pair of corresponding bits in the two regex states.
If \isa{None} is returned for any pair, then the function returns \isa{None} for the entire regex state.
Note that the lengths of the two regex states must be the same (i.e., equal to $n$, the number of atomic propositions), and this operation returns \isa{None} if they are not.
Then, we again lift \isa{WEST\_and\_state} to operate on two trace regexes in the function \isa{WEST\_and\_trace} by applying \isa{WEST\_and\_state} to each pair of corresponding regex states in the two trace regexes, returning \isa{None} if any of the calls to \isa{WEST\_and\_state} returns \isa{None}.
The input trace regexes are allowed to have different lengths, and the shorter trace regex is treated as if the missing regex states are all \isa{S}, following \cite[Definition 4, Pad]{DBLP:conf/ifm/ElwingGSTWR23}.
The full formal definitions can be found in our formalization.

To establish the correctness of \isa{WEST\_and}, we prove the following lemma:
\begin{mdframed}[backgroundcolor=black!10,linecolor=black!10]
\begin{isa}
    \WESTandcorrect
\end{isa}
\end{mdframed}
This shows that for input WEST regexes \isa{L1} and \isa{L2}, both \isa{L1} and \isa{L2} match trace \isasympi\ iff the \isa{WEST\_and} of \isa{L1} and \isa{L2} matches \isasympi.
In other words, the set of traces that the \isa{WEST\_and} of \isa{L1} and \isa{L2} matches is exactly the intersection between the set of traces that \isa{L1} matches and the set of traces that \isa{L2} matches.
The assumptions on \isa{L1} and \isa{L2} are well-definedness conditions that ensure all state regexes have length $n$ (the number of atomic propositions), as required by \isa{WEST\_and\_state}.

To keep the sizes of WEST regexes small, WEST implements an additional simplification step which collects together related trace regexes.
If two trace regexes differ only by a single bit, then they may be combined into one trace regex where the differing bit is \isa{S}. 
For example, fixing the number of atomic propositions to $n=2$, the WEST regex
\isa{[[[0,0],[0,1]], [[0,0],[0,0]], [[0,1],[0,S]]]} may first be reduced (by combining the first two trace regexes) to \isa{[[[0,0],[0,S]], [[0,1],[0,S]]]}, and then to \isa{[[[0,S],[0,S]]]}.
This is crucial for improving the tool performance, as it helps to mitigate blowup in the length of the list of trace regexes during the \texttt{WEST\_and} and \texttt{WEST\_or} operations \cite[Section 4]{DBLP:conf/ifm/ElwingGSTWR23}. 

The underlying idea is straightforward: greedily simplify pairs of regexes until no more pairs can be simplified; we implement this in the \isa{WEST\_simp} function.
It is crucial that the simplification step does not change the set of traces that a WEST regex matches.
The following lemma shows that, for a well-defined WEST regex \isa{L}, a trace \isasympi\ matches \isa{L} iff \isasympi\ matches the simplification of \isa{L}:
\begin{mdframed}[backgroundcolor=black!10,linecolor=black!10]
    \begin{isa}
        \WESTsimpcorrect
    \end{isa}
\end{mdframed}
Finally, we define the functions \isa{WEST\_and\_simp} and \isa{WEST\_or\_simp} by passing the output of \isa{WEST\_and} and \isa{WEST\_or} (respectively) to \isa{WEST\_simp}.
The correctness of \isa{WEST\_and\_simp} and \isa{WEST\_or\_simp} follows directly from the correctness results for \isa{WEST\_and}, \isa{WEST\_or}, and \isa{WEST\_simp}.

\subsection{Temporal Operators}\label{sec:temporal}
Our formalization of the temporal operators in the WEST algorithm uses the \isa{WEST\_and} and \isa{WEST\_or} operators.
It also makes use of an operation to shift regular expressions to later timesteps, which we call \isa{shift}.
Though the source material never explicitly defines this \isa{shift} operation, it uses it implicitly and defines an analogous operation \cite[Definition 5]{DBLP:conf/ifm/ElwingGSTWR23}. 
We formalize \isa{shift} 
as follows:

\begin{mdframed}[backgroundcolor=black!10,linecolor=black!10]
\begin{isa}
    \shift
\end{isa}
\end{mdframed}

Here, we refer to a state regex of all \isa{S}'s as an \textit{arbitrary state}, and we refer to a trace regex of all arbitrary states as an \textit{arbitrary trace} \cite[Section 6]{DBLP:conf/ifm/ElwingGSTWR23}. 
In this snippet, \isa{arbitrary\_trace n t} constructs an arbitrary trace regex containing \isa{t} arbitrary states of length \isa{n}.
Then, \isa{shift} takes as input a WEST regex \isa{L}, and appends an arbitrary trace of \isa{t} arbitrary states to the front of each trace regex in \isa{L}.
As intuitively named, \isa{shift} shifts all trace regexes in \isa{L} by \isa{t} timesteps.

For example, fixing the number of atomic propositions at $n = 2$, the WEST regex \isa{L = [[[1,1]], [[0,0], [0,0]]]} captures that either $p_0$ and $p_1$ both need to be true at timestep $0$, or $p_0$ and $p_1$ both need to be false at timesteps $0$ and $1$.
If instead we want to delay this behavior for $p_0$ and $p_1$ by $3$ timesteps, we can compute \isa{shift L 2 3}, which returns \isa{[[[S,S],[S,S],[S,S],[1,1]], [[S,}
\isa{S],[S,S],[S,S],[0,0],[0,0]]]}.
The following lemma formalizes the connection between the \isa{shift} operation for WEST regexes and the suffix of a trace:

\begin{mdframed}[backgroundcolor=black!10,linecolor=black!10]
\begin{isa}
    \padmatchproperty
\end{isa}
\end{mdframed}
More precisely, \isa{shift\_match\_property} establishes that a sufficiently long trace \isasympi\ matches a WEST regex \isa{L} shifted by \isa{t} timesteps iff the suffix of \isasympi\ with \isa{t}\ states removed, denoted \isa{drop t \isasympi}, matches \isa{L}.

Now, we demonstrate how the temporal operators are built on top of the core WEST operators.
We provide for an example \isa{WEST\_global}, defined as follows:
\begin{mdframed}[backgroundcolor=black!10,linecolor=black!10]
    \begin{isa}
        \WESTglobal
    \end{isa}
\end{mdframed}

\isa{WEST\_global} takes as input a WEST regex \isa{L}, lower and upper interval bounds \isa{a} and \isa{b}, and the number of atomic propositions \isa{n}. 
\isa{WEST\_global} then uses the \isa{shift} operation to shift the input regex \isa{L} by \isa{b} timesteps, and computes the \isa{WEST\_and} of the shifted \isa{L} and \isa{WEST\_global} with \isa{b-1}. 
Intuitively, \isa{L} captures a set of traces specifying some behavior at timestep $0$, and the successive \isa{shift} and \isa{WEST\_and} operations ensures that \isa{L}'s behavior happens at all timesteps between \isa{a} and \isa{b}.
The remaining temporal operators are defined in a similar manner, using \isa{shift} and the core WEST operators.

We establish the correctness of the \isa{WEST\_global} operator as follows:
\begin{mdframed}[backgroundcolor=black!10,linecolor=black!10]
    \begin{isa}
        \WESTglobalcorrect
    \end{isa}
\end{mdframed}
This lemma says that for a WEST regex \isa{L} over \isa{n} variables (assumption \isa{L\_vars}) that captures the semantics of an MLTL formula \isasymphi\ of at most \isa{n} variables (assumption \isa{semantics\_\isasymphi} and \isa{\isasymphi\_vars}), and a trace \isasympi\ of sufficient length, 
\isa{WEST\_global \isasymphi\ a b n} matches \isasympi\ iff \isasympi\ satisfies the semantics of \isa{Global\_mltl \isasymphi\ a b} (representing the formula $\texttt{G}_{[a,b]} \phi$). 

Likewise, each of the remaining temporal operators has a correctness lemma that establishes the connection between the WEST regex it computes and its corresponding temporal operator. 
The correctness lemmas for the temporal operators totaled about 850 lines of code.

\subsection{Top-Level Algorithm and Correctness}\label{sec:toplevel}

The WEST algorithm takes as input an MLTL formula $\phi$ in negation normal form (NNF) and recursively computes the WEST regex representing the set of traces with computation length \isasymphi\ that satisfy the formula.
The existing Isabelle/HOL MLTL library \cite{KWSR24} already formalizes the computation length\footnote{This is also known as the \textit{worst-case propagation delay} in the context of runtime verification \cite{KZJZR20,ZADJR23}.}
of $\phi$, denoted \texttt{complen}($\phi$), which intuitively measures how much time is needed to decide the satisfiability of $\phi$ \cite{DBLP:conf/ifm/ElwingGSTWR23,KZJZR20,KWSR24}.

We formalize the WEST algorithm in the function \isa{WEST\_reg} as follows:
\begin{mdframed}[backgroundcolor=black!10,linecolor=black!10]
    \begin{isa}
        \WESTreg
    \end{isa}
\end{mdframed}
Although input formulas to the WEST algorithm must be in NNF, we allow formulas of all shapes as input and apply the \isa{convert\_nnf} function from existing work \cite{KWSR24} to transform the input formula to NNF.
The resultant NNF formula \isa{nnf\_\isasymphi} and the number of atomic propositions, computed as \isa{WEST\_num\_vars\ \isasymphi}, are then passed to the auxiliary function \isa{WEST\_reg\_aux}. 
This auxiliary function takes two inputs (a \isa{nat\ mltl} formula \isasymphi\ and a natural number \isa{n} for the number of atomic propositions) and cases on the structure of \isasymphi\ to apply the appropriate core operators and return a WEST regex.

We consider here a few representative cases: \isa{True}, \isa{Prop\_mltl}, \isa{And\_mltl}, and \isa{Global}\isa{\_mltl} (corresponding to the cases of \texttt{True}, an atomic proposition, a conjunction, and the global operator). 
Mathematically, these cases are defined in the source material as follows \cite{DBLP:conf/ifm/ElwingGSTWR23}: $\texttt{reg}(\texttt{True}) = S^n$, $\texttt{reg}(p_k) = S^{k}1S^{n-k-1}$, and $\texttt{reg}(\phi \land \psi) = \texttt{reg}(\phi) \land \texttt{reg}(\psi)$.  The global operation, $\texttt{reg}(\texttt{G}_{[a,b]} \phi)$ computes (recursively) the \texttt{WEST\_and} of $\texttt{reg}(\phi)$ shifted by $i$ timesteps for all $i$ with $a \leq i \leq b$ (note this is essentially what \isa{WEST\_global} computes).
In Isabelle/HOL, we have:

\begin{mdframed}[backgroundcolor=black!10,linecolor=black!10]
    \isa{
        \ WEST{\isacharunderscore}{\kern0pt}reg{\isacharunderscore}{\kern0pt}aux{\isacharcolon}{\kern0pt}{\isacharcolon}{\kern0pt}\ {\isachardoublequoteopen}{\isacharparenleft}{\kern0pt}nat{\isacharparenright}{\kern0pt}\ mltl\ {\isasymRightarrow}\ nat\ {\isasymRightarrow}\ WEST{\isacharunderscore}{\kern0pt}regex{\isachardoublequoteclose}\ \isanewline
        \isakeyword{where}\ {\isachardoublequoteopen}WEST{\isacharunderscore}{\kern0pt}reg{\isacharunderscore}{\kern0pt}aux\ True{\isacharunderscore}{\kern0pt}mltl\ n\ {\isacharequal}{\kern0pt}\ {\isacharbrackleft}{\kern0pt}{\isacharbrackleft}{\kern0pt}{\isacharparenleft}{\kern0pt}map\ {\isacharparenleft}{\kern0pt}{\isasymlambda}\ j{\isachardot}{\kern0pt}\ S{\isacharparenright}{\kern0pt}\ {\isacharbrackleft}{\kern0pt}{\isadigit{0}}\ {\isachardot}{\kern0pt}{\isachardot}{\kern0pt}{\isacharless}{\kern0pt}\ n{\isacharbrackright}{\kern0pt}{\isacharparenright}{\kern0pt}{\isacharbrackright}{\kern0pt}{\isacharbrackright}{\kern0pt}{\isachardoublequoteclose}\isanewline
        {\isacharbar}{\kern0pt}\ {\isachardoublequoteopen}WEST{\isacharunderscore}{\kern0pt}reg{\isacharunderscore}{\kern0pt}aux\ {\isacharparenleft}{\kern0pt}Prop{\isacharunderscore}{\kern0pt}mltl\ p{\isacharparenright}{\kern0pt}\ n\ {\isacharequal}{\kern0pt}\ \isanewline
        \ \ \ {\isacharbrackleft}{\kern0pt}{\isacharbrackleft}{\kern0pt}{\isacharparenleft}{\kern0pt}map\ {\isacharparenleft}{\kern0pt}{\isasymlambda}j{\isachardot}{\kern0pt}\ {\isacharparenleft}{\kern0pt}if\ {\isacharparenleft}{\kern0pt}p{\isacharequal}{\kern0pt}j{\isacharparenright}{\kern0pt}\ then\ One\ else\ S{\isacharparenright}{\kern0pt}{\isacharparenright}{\kern0pt}\ {\isacharbrackleft}{\kern0pt}{\isadigit{0}}\ {\isachardot}{\kern0pt}{\isachardot}{\kern0pt}{\isacharless}{\kern0pt}\ n{\isacharbrackright}{\kern0pt}{\isacharparenright}{\kern0pt}{\isacharbrackright}{\kern0pt}{\isacharbrackright}{\kern0pt}{\isachardoublequoteclose}\isanewline
        {\isacharbar}{\kern0pt}\ {\isachardoublequoteopen}WEST{\isacharunderscore}{\kern0pt}reg{\isacharunderscore}{\kern0pt}aux\ {\isacharparenleft}{\kern0pt}And{\isacharunderscore}{\kern0pt}mltl\ {\isasymphi}\ {\isasympsi}{\isacharparenright}{\kern0pt}\ n\ {\isacharequal}{\kern0pt}\ {\isacharparenleft}{\kern0pt}WEST{\isacharunderscore}{\kern0pt}and{\isacharunderscore}{\kern0pt}simp\ \isanewline
        \ \ \ {\isacharparenleft}{\kern0pt}WEST{\isacharunderscore}{\kern0pt}reg{\isacharunderscore}{\kern0pt}aux\ {\isasymphi}\ n{\isacharparenright}{\kern0pt}\ {\isacharparenleft}{\kern0pt}WEST{\isacharunderscore}{\kern0pt}reg{\isacharunderscore}{\kern0pt}aux\ {\isasympsi}\ n{\isacharparenright}{\kern0pt}\ n{\isacharparenright}{\kern0pt}{\isachardoublequoteclose}\isanewline
        {\isacharbar}{\kern0pt}\ {\isachardoublequoteopen}WEST{\isacharunderscore}{\kern0pt}reg{\isacharunderscore}{\kern0pt}aux\ {\isacharparenleft}{\kern0pt}Global{\isacharunderscore}{\kern0pt}mltl\ {\isasymphi}\ a\ b{\isacharparenright}{\kern0pt}\ n\ {\isacharequal}{\kern0pt}\ \isanewline
        \ \ \ WEST{\isacharunderscore}{\kern0pt}global\ {\isacharparenleft}{\kern0pt}WEST{\isacharunderscore}{\kern0pt}reg{\isacharunderscore}{\kern0pt}aux\ {\isasymphi}\ n{\isacharparenright}{\kern0pt}\ a\ b\ n{\isachardoublequoteclose}
    }
\end{mdframed}
Here, \isa{map f L} applies a function \isa{f} on every element of a list \isa{L}, so the base case for \isa{True\_mltl} creates a WEST regex containing a trace regex of all \isa{S}. 
In the case \isa{Prop\_mltl p}, the map function takes as input \isa{j} and returns \isa{One} if the propositional variable \isa{p} equals the index \isa{j}, and otherwise \isa{S}.
In \isa{And\_mltl}, we directly call the \isa{WEST\_and} operator; likewise in \isa{Global\_mltl}.

\paragraph{Top-Level Correctness.} 
A central contribution of our work is proving (and even generalizing slightly) the correctness of the \isa{WEST\_reg\_aux} function and elucidating many of the details omitted in the original proof of correctness.
Theorem $2$ in the source material states the correctness result as follows: for a MLTL formula $\phi$ in negation normal form, a trace $\pi$ with length $\texttt{complen}(\phi)$ satisfies $\phi$ iff $\pi$ matches $\texttt{reg}(\phi)$ \cite{DBLP:conf/ifm/ElwingGSTWR23}.
We formalize this in the theorem \isa{WEST\_reg\_aux\_correct}:

\begin{mdframed}[backgroundcolor=black!10,linecolor=black!10]
    \begin{isa}
        \WESTregauxcorrect
    \end{isa}
\end{mdframed}

This theorem states that for MLTL formula \isasymphi\ in NNF (assumption \isa{is\_nnf}) with at most \isa{n} variables (assumption \isa{\isasymphi\_nv}) and well-defined interval bounds (assumption \isa{intervals\_welldef \isasymphi}) and a trace \isasympi\ of length at least \texttt{complen}(\isasymphi)\ (assumption \isa{\isasympi\_long\_enough}), the trace \isasympi\ satisfies \isasymphi\ iff the trace \isasympi\ matches the WEST regex computed by \isa{WEST\_reg\_aux \isasymphi\ n}.
Here, the functions \isa{convert\_nnf}, \isa{complen\_mltl}, and \isa{intervals\_welldef} are from existing work \cite{KWSR24}.
The \isa{\isasymphi\_nv} is an implicit assumption in the source material, which globally fixes the number of atomic propositions.\footnote{Note that we crucially assume that the number of variables of \isasymphi\ is $\leq n$ instead of $=n$ in order to satisfy the inductive hypothesis in our (inductive) proof.}
We slightly generalize the original correctness result, as our formal result holds for all traces of length at least the computation length of \isasymphi\, rather than just the traces of length equal to the computation length of \isasymphi. 

We prove this by structural induction on the input formula \isasymphi. 
The \isa{is\_nnf} assumption allows us to use the custom induction rule \isa{nnf\_induct} from prior work \cite{KWSR24}, simplifying the induction proof.
The base cases are straightforward, and the inductive cases are proven by applying the inductive hypothesis on the subformulas and using the correctness lemmas for the core WEST operators.
For instance, for input formula \isa{\isasymphi\ = Global\_mltl \isasympsi\ a b} (which is $\texttt{G}_{[a, b]} \psi$), the inductive hypothesis gives us that the trace \isasympi\ satisfies \isasympsi\ iff the WEST regex \isa{L} computed by \isa{WEST\_reg\_aux \isasympsi\ n} matches \isasympi.
Next, in order to apply the correctness result of the \isa{WEST\_global} operator, we need to show that \isa{L} is a WEST regex over \isa{n} atomic propositions (i.e., each state regex in each trace regex in \isa{L} is of length \isa{n}).
For this, we prove the lemma \isa{WEST\_reg\_aux\_num\_vars}:

\begin{mdframed}[backgroundcolor=black!10,linecolor=black!10]
    \begin{isa}
        \WESTregauxnv
    \end{isa}
\end{mdframed}
This lemma states that, for a formula \isasymphi\ in NNF with at most \isa{n} atomic propositions, the WEST regex computed by \isa{WEST\_reg\_aux \isasymphi\ n} is a WEST regex over \isa{n} atomic propositions.
With this, we can apply the correctness result of the \isa{WEST\_global} operator on \isa{L} and complete the proof of the \isa{Global\_mltl} case.

Finally, we present the top-level correctness result for the WEST algorithm:

\begin{mdframed}[backgroundcolor=black!10,linecolor=black!10]
    \begin{isa}
        \WESTregcorrect
    \end{isa}
\end{mdframed}
This theorem states that for any MLTL formula \isasymphi\ with well-defined interval bounds \cite{KWSR24} and any trace \isasympi\ of length at least the computation length of \isasymphi, \isasympi\ satisfies \isasymphi\ iff the WEST regex \isa{WEST\_reg \isasymphi} matches \isasympi.
The correctness of the top-level WEST algorithm took about 600 LOC in Isabelle/HOL compared to the 60 or so lines of proof sketches in the source material \cite[Appendix III]{DBLP:conf/ifm/ElwingGSTWR23}.\footnote{The results leading up to this top-level theorem required an additional $\approx$ 5300 LOC.}

\section{Formalization Insights}
\label{sec:insights}
Retrospectively viewing our formalization at a high level, we highlight a few notable points.
First, our modular definitions did considerably streamline our correctness proofs.
Many proofs have relatively similar structures, which helped guide the formalization at a high level.
However, we also found that relatively short proofs in the source material became lengthy in the formalization, in part because they often split into many subcases.
For example, the notion of a WEST regex matching a trace is intuitively simple, but the formalization used several helper functions.
As another example, the proof of \isa{WEST\_and\_correct} is approximately 15 lines of a proof sketch in the source material \cite[Theorem 4]{DBLP:conf/ifm/ElwingGSTWR23}.
However, our formal development took approximately 1800 LOC to state and prove this result level by level, starting from the correctness of the \texttt{and} operation on state regexes, then on trace regexes, and finally on WEST regexes.
Although these proofs had structural similarities, subtle differences between the operators complicated the low-level details of the proofs; for instance, the option types of \isa{WEST\_and\_state} required careful analysis in the correctness proofs.

Second, our formalization makes \textit{all} details explicit, including details omitted in the source material.
Many of our formal proofs are by induction; setting up the ``right'' inductive structure in a formal setting requires careful analysis that is often glossed over in source material.
For instance, the top-level correctness theorem required making a mathematically implicit assumption on \isa{num\_vars} explicit.
Setting up this assumption in the wrong way leads to an ineffective inductive structure. 
As another example, in the proof of \isa{WEST\_simp\_correct}, we perform a tricky induction on the difference between the length of the input WEST regex and the output simplified WEST regex.
Additionally, we are required to prove that all functions terminate.
For many functions, Isabelle/HOL proves this automatically \cite{KraussThesis}, but we occasionally ran into cases where we had to explicitly construct a measure to prove termination.
For example, the \isa{WEST\_reg\_aux} function and the \isa{WEST\_simp} function required such manual termination proofs.
Intuitively, \isa{WEST\_reg\_aux} recurses on all subformulas in NNF, converting subformulas to NNF as necessary; accordingly, we use a termination measure that is similar to the number of nodes in the abstract syntax tree (AST) of the formula, but weighs nodes that are not in NNF more heavily. 
This allows us to prove that \isa{WEST\_reg\_aux} terminates, as this measure strictly decreases on every recursive call. 
Further, for \isa{WEST\_simp}, the length of the input list is not strictly decreasing, but the list of candidate pairs for simplification will be exhausted at some point, so we use a measure that combines the length of the input list with this. 

Overall, integrating \isa{WEST\_simp} into our formalization was rather involved.
Our initial formalization did not include \isa{WEST\_simp}, but we ultimately realized that it is crucial for speed and thus also important for tool validation.
While the modular nature of our formalization easily allowed us to add in this function to the algorithm, its correctness proofs were intricate.
Similarly to \isa{WEST\_and}, we proved the correctness of \isa{WEST\_simp} level by level, totaling around 1300 LOC.

As a final interesting point, we found during our tool validation that \isa{WEST\_reg} and the (unverified) WEST tool sometimes produce trace regexes that differ only by a string of \isa{S}'s at the end.
In such cases, because these trace regexes have different length, our equivalence checking methods spuriously identify a mismatch.
The WEST tool always produces trace regexes that have the same length as the computation length of the input formula, while \isa{WEST\_reg} does not.\footnote{This is because we implicitly treat shorter trace regexes to have all \isa{S}'s at the end (recall our discussion of the \isa{WEST\_and\_trace} operator in \rref{sec:core}).}
To account for this, we define a function \isa{simp\_pad\_WEST\_reg} which pads trace regexes to this computation length (and then simplifies).
We extend the top-level correctness theorem from \isa{WEST\_reg} to \isa{simp\_pad\_WEST\_reg}; in our experiments, we work with \isa{simp\_pad\_WEST\_reg} so as to eliminate these spurious mismatches.
 
\section{Experiments}
\label{sec:experiments}
The functions \isa{simp\_pad\_WEST\_reg} and \isa{naive\_equivalence} are executable in Isabelle/HOL, and we use Isabelle/HOL's code generator \cite{DBLP:phd/de/Haftmann2009} to export these functions to Haskell.\footnote{Note that, although Isabelle/HOL's code generator is not yet fully verified, exporting a formalized function is more trustworthy than simply coding a function. Additionally, some work has considered verifying Isabelle's code generator \cite{DBLP:conf/esop/HupelN18}.}
We choose Haskell both to facilitate our experimental setup and because the GHC compiler \cite{GHC} produces reasonably fast native machine code.
We use our code export to validate two versions of the WEST tool---the initial version of WEST \cite{DBLP:conf/ifm/ElwingGSTWR23}, and also a more recent version that has been highly optimized \cite{WEST2024tool}.
We also compare the different implementations for speed.
We run all of our experiments in WSL2 on a Windows machine with an 11th generation Intel Core i7 processor and 32GB of RAM.
We use an unverified parsing script to transform input MLTL formulas into the format required by our code export.\footnote{There has been some recent work \cite{LL1ParserAFP} on improving support for verified parsing in Isabelle/HOL, so verifying this parsing step might be an interesting future direction.} 

\paragraph{Previous Validation Efforts.}
The most recent (and fastest) version of WEST was validated against several MLTL tools \cite{WEST2024tool}: 
\textcircled{1} the original version of WEST \cite{DBLP:conf/ifm/ElwingGSTWR23}, 
\textcircled{2} the runtime verification engine R2U2 \cite{RS17,KZJZR20,JJKRZ23}
\textcircled{3} a direct C++ implementation of MLTL semantics \cite{WEST2024tool}, and 
\textcircled{4} translating MLTL formulas to propositional logic \cite{HJRW23} and applying a BDD based AllSAT solver.
The validation works by analyzing, for each formula in the test suite, whether the trace set of regexes produced by WEST is equivalent to the set of satisfying traces produced by other tools.
The equivalence checking is a crucial step performed between outputs that can be in different formats (depending on the output format of each tool).
The test suite of $1662$ MLTL formulas was designed to capture every possible combination of MLTL operators \cite{DBLP:conf/ifm/ElwingGSTWR23}.

\subsection{Verified Equivalence Checking}
Our tool validation is set up to check the outputs of our verified implementation of WEST against the two existing implementations.
For this, we need to be able to check equivalences between WEST regexes.
It is not always enough to merely check set equality, as implementation differences can lead to different (but logically equivalent) outputs.
For instance, the two WEST regexes $[[[S,S]]]$ and $[[[S,1]], [[1,S]], [[0,0]]]$ are equivalent, but \isa{WEST\_simp} does not simplify the second into the first. 
The order in which \isa{WEST\_simp} simplifies pairs of trace regexes within a WEST regex is what causes these differences. 

Developing a fully verified and optimized equivalence checking algorithm is out of scope of our work, but we still wanted a lightweight trustworthy implementation of regex equivalence checking.
Accordingly, we formalize a naive equivalence checking function for WEST regexes, called \isa{naive\_equivalence}.
This function works by explicitly enumerating all the trace regexes that each WEST regex produces and then checking set equality.

We then prove the experimentally relevant direction of correctness: If two WEST regexes are equivalent under our (executable) naive equivalence checking function, then they are indeed equivalent under the (non-executable) mathematical definition.
Formally, we have the following lemma:
\begin{mdframed}[backgroundcolor=black!10,linecolor=black!10]
    \begin{isa}
        \regexequivalencecorrect
    \end{isa}
\end{mdframed}
The proof was approximately 1150 lines of code.
Although establishing both directions of equivalence here (i.e, $\longleftrightarrow$ instead of $\longrightarrow$) is theoretically desirable, the direction we verify is the experimentally significant one, since we encounter no instances where \isa{naive\_equivalence} failed in our test suite.
More specifically, \isa{naive\_equivalence} holds on all but 4 of the 1662 input formulas and times out (after 4 hours) on the remaining 4 formulas. 
Often the outputs are identical; for example, the Isabelle implementation and the optimized WEST tool produced identical WEST regexes on 1547 of the formulas.
We additionally ran the previous (unverified) equivalence checking procedure, which succeeded on all of the formulas.
Collectively, these results establish strong confidence in the correctness of the (unverified) WEST tools \cite{DBLP:conf/ifm/ElwingGSTWR23,WEST2024tool}.

\subsection{Speed Comparison}
The original C++ version of WEST \cite{DBLP:conf/ifm/ElwingGSTWR23} performed string-based operations, and the optimized version of WEST takes advantage of highly parallelized computations by using bitsets \cite{WEST2024tool}.
Although fast performance is not our primary goal, preliminary experiments demonstrate how our formalized code compares to the two unverified versions of WEST.
Overall, we find that the optimized version of WEST is fast (as expected).
\begin{figure}[h]
\begin{minipage}{0.50\textwidth}
    \centering
    \includegraphics[width=1\linewidth]{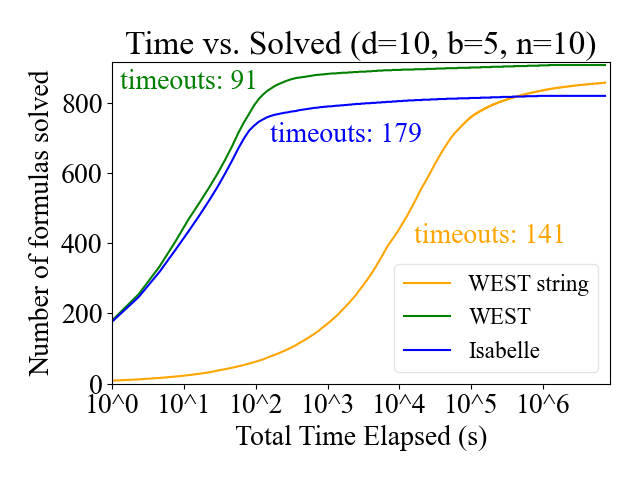}
\end{minipage}
\hfill
\begin{minipage}{0.50\textwidth}
    \centering
    \includegraphics[width=1\linewidth]{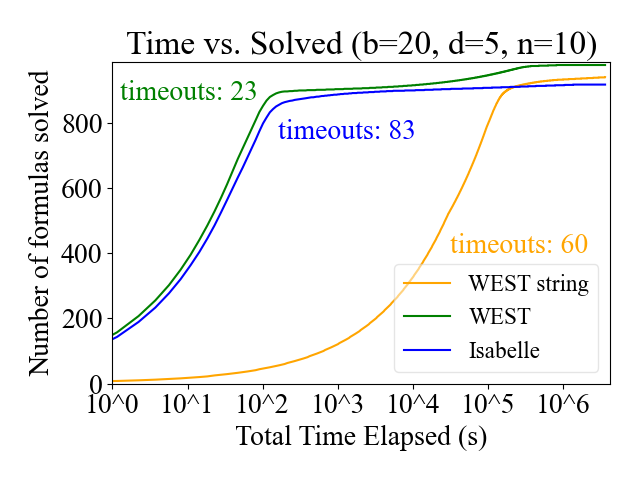}
\end{minipage}
\caption{Two cactus plots, each comparing the three WEST implementations on $1000$ random formulas of varying nesting depth $d$, interval bounds $b$, and number of atomic propositions $n$.
The number of total solved instances is shown on the y-axis, and the cumulative time taken is shown on the x-axis, with the number of timeouts labeled. 
}
\label{fig:cactus}
\end{figure}
Our Isabelle implementation also performs quite respectably; it is, in aggregate, close in performance to the optimized version of WEST.
We perform extensive experiments to compare the performance of the three tools on large randomly generated benchmark sets. 
We use a script to generate random MLTL formulas \cite{WEST2024tool}, varying the parameters of the maximum depth and the maximum interval time bounds.
Our results are in \rref{fig:cactus}.
As the primary focus of our work is tool validation, we do not envision our contribution as replacing the WEST tool, but its relative efficiency is encouraging nonetheless. 

\begin{figure}[!htb]
\begin{minipage}{0.48\textwidth}
    \centering
    \includegraphics[width=1\linewidth]{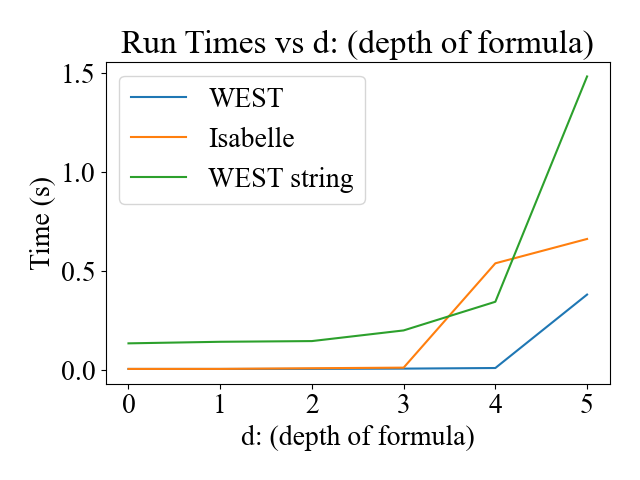}
\end{minipage}
\hfill
\begin{minipage}{0.48\textwidth}
    \centering
    \includegraphics[width=1\linewidth]{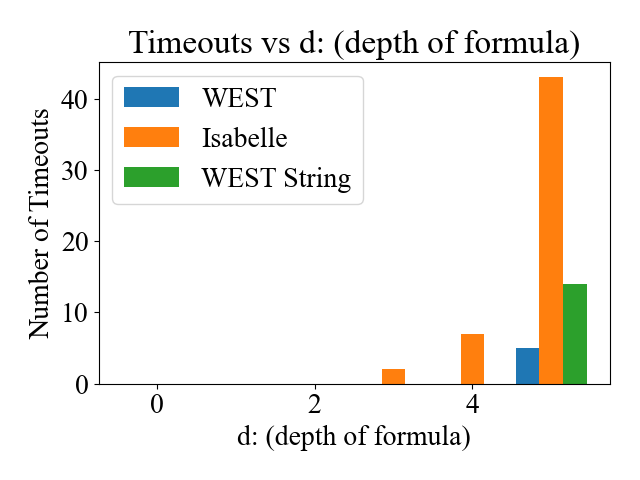}
\end{minipage}
\caption{Results for $n=5$, $b=2$, and varying values of $d$ from $0$ to $5$, with a batch size of 300 formulas per value of $d$.
The Isabelle implementation is faster than the unoptimized WEST tool on most values of $d$, but times out on many formulas for $d=5$. 
}
\label{fig:experiments_easyd}
\end{figure}

Additionally, we did find that, on individual examples, our code export has somewhat unpredictable behavior (whereas the optimized version of WEST appears to be uniformly fast), and our code export seems to incur timeouts more frequently than the unverified WEST implementations.
For example, in \rref{fig:experiments_easyd}, we evaluate the speed of the three tools based on varying values of $d$, the depth of the formula, while fixing the number of atomic propositions at $n=5$ and maximum interval bound to $b=2$. 
Here, we observe that the Isabelle implementation begins to time out much more frequently than the other two tools when $d=4$ and $d=5$. 

Additional results, including aggregate cactus plots on easier but larger test suites, an extension of \rref{fig:experiments_easyd} on higher values of formula depth $d$, and experiments where we vary the value of maximum interval bound $b$ (instead of $d$), can be found in \rref{app:experiments}.

\section{Conclusion}
\label{sec:conclusion}

Our work produces a third, open-source, freely available implementation of the WEST algorithm, this time \textit{formally} verified \cite{Mission_Time_LTL_to_Regular_Expression-AFP}.
Given the popularity of MLTL as a formal specification language for safety-critical applications \cite{KZJZR20,DBR21,LJBHCLR22,DRB22,AJR22}, verifying significant algorithms like WEST, which facilitates MLTL specification, is well-justified.
We build on an existing formalization of MLTL in Isabelle/HOL \cite{KWSR24} to further develop the library of verified MLTL algorithms and properties, which could help facilitate future verified developments in this space.
Our development validates the existing (unverified) WEST tool \cite{DBLP:conf/ifm/ElwingGSTWR23,WEST2024tool} on benchmarks from the literature, bringing us a step closer to validating other MLTL tools like R2U2 \cite{RRS14,JJKRZ23}. 
Though our primary focus was not on speed, the aggregate performance of our Isabelle-generated code is promising, and optimizing our formalization could be interesting future work.
It would be particularly beneficial to further optimize (and verify the reverse direction of) our naive WEST regex equivalence checking, possibly using existing work \cite{DBLP:journals/jar/KraussN12} which verifies regex equivalence checking in a general setting.
Verified parsing (to transform input formulas into the syntax required by our code export) would also be welcome.
Additionally, a deeper analysis of the performance of the WEST tools and of our verified code on different classes of benchmarks could inform future verified tool generation efforts. 
For example, it would be interesting to experimentally compare a code export to some of the other languages supported by Isabelle/HOL, like SML and OCaml, to see if a different target language could help avoid timeouts.
Importantly, our formalization of MLTL rewriting, equivalence checking, and regular expression manipulation could serve as a basis for formalizing similar utilities in logics like MTL and STL that extend MLTL. 

\begin{credits}
\subsubsection{\ackname} Thanks to NSF CAREER Award CNS-1552934, NSF CCRI-2016592, and GRFP-2024364991 for supporting this work.   
We thank the annonymous TACAS reviewers as well as Alec Rosentrater and Laura Gamboa Guzman for their helpful feedback on the paper, and the TACAS artifact evaluators for their time. 
 \end{credits}

%
%
%
\bibliographystyle{splncs04}
\bibliography{tacas2025}

\begin{thebibliography}{10}
\providecommand{\url}[1]{\texttt{#1}}
\providecommand{\urlprefix}{URL }
\providecommand{\doi}[1]{https://doi.org/#1}

\bibitem{AH90}
Alur, R., Henzinger, T.A.: {R}eal-time {L}ogics: {C}omplexity and
  {E}xpressiveness. In: LICS. pp. 390--401. IEEE (1990)

\bibitem{DBLP:conf/podc/AlurFH91}
Alur, R., Feder, T., Henzinger, T.A.: {The Benefits of Relaxing Punctuality}.
  In: Logrippo, L. (ed.) Proceedings of the Tenth Annual {ACM} Symposium on
  Principles of Distributed Computing, Montreal, Quebec, Canada, August 19-21,
  1991. pp. 139--152. {ACM} (1991). \doi{10.1145/112600.112613},
  \url{https://doi.org/10.1145/112600.112613}

\bibitem{LTL3_Semantics-AFP}
Amjad, R., van Glabbeek, R., O'Connor, L.: Definitive set semantics for {LTL3}.
  Archive of Formal Proofs  (August 2024),
  \url{https://isa-afp.org/entries/LTL3\_Semantics.html}, Formal proof
  development

\bibitem{Mav22}
{Anastasia Mavridou}: {Capturing and Analyzing Requirements with FRET}.
  Presentation, nasa formal methods symposium,
  \url{https://github.com/NASA-SW-VnV/fret}, National Aeronautics and Space
  Agency, Pasadena, California, USA (May 2022)

\bibitem{AJR22}
Aurandt, A., Jones, P., Rozier, K.Y.: {Runtime Verification Triggers Real-time,
  Autonomous Fault Recovery on the CySat-I}. In: {Proceedings of the 14th NASA
  Formal Methods Symposium (NFM 2022)}. Lecture Notes in Computer Science
  (LNCS), vol. 13260. Springer, Cham, Caltech, California, USA (May 2022).
  \doi{10.1007/978-3-031-06773-0\_45}

\bibitem{DBLP:conf/types/Ballarin03}
Ballarin, C.: Locales and locale expressions in {Isabelle/Isar}. In: Berardi,
  S., Coppo, M., Damiani, F. (eds.) TYPES. LNCS, vol.~3085, pp. 34--50.
  Springer (2003). \doi{10.1007/978-3-540-24849-1\_3},
  \url{https://doi.org/10.1007/978-3-540-24849-1\_3}

\bibitem{DBLP:journals/jar/Ballarin14}
Ballarin, C.: Locales: {A} module system for mathematical theories. J. Autom.
  Reason.  \textbf{52}(2),  123--153 (2014). \doi{10.1007/S10817-013-9284-7},
  \url{https://doi.org/10.1007/s10817-013-9284-7}

\bibitem{DBLP:conf/ictac/BasinDHHMKKMRST22}
Basin, D.A., Dardinier, T., Hauser, N., Heimes, L., y~Munive, J.J.H., Kaletsch,
  N., Krstic, S., Marsicano, E., Raszyk, M., Schneider, J., Tirore, D.L.,
  Traytel, D., Zingg, S.: Veri{M}on: {A} formally verified monitoring tool. In:
  Seidl, H., Liu, Z., Pasareanu, C.S. (eds.) ICTAC. LNCS, vol. 13572, pp.~1--6.
  Springer (2022). \doi{10.1007/978-3-031-17715-6\_1},
  \url{https://doi.org/10.1007/978-3-031-17715-6\_1}

\bibitem{DBLP:conf/cav/CavadaCDGMMMRT14}
Cavada, R., Cimatti, A., Dorigatti, M., Griggio, A., Mariotti, A., Micheli, A.,
  Mover, S., Roveri, M., Tonetta, S.: The {nuXmv} symbolic model checker. In:
  Biere, A., Bloem, R. (eds.) CAV. LNCS, vol.~8559, pp. 334--342. Springer
  (2014). \doi{10.1007/978-3-319-08867-9\_22},
  \url{https://doi.org/10.1007/978-3-319-08867-9\_22}

\bibitem{OnlineMTL_Coq}
Chattopadhyay, A., Mamouras, K.: {A Verified Online Monitor for Metric Temporal
  Logic with Quantitative Semantics}. In: Runtime Verification: 20th
  International Conference, RV 2020, Los Angeles, CA, USA, October 6–9, 2020,
  Proceedings. p. 383–403. Springer-Verlag, Berlin, Heidelberg (2020).
  \doi{10.1007/978-3-030-60508-7\_21},
  \url{https://doi.org/10.1007/978-3-030-60508-7\_21}

\bibitem{MTL_monitor_coq}
Chattopadhyay, A., Mamouras, K.: A verified online monitor for metric temporal
  logic with quantitative semantics. In: Deshmukh, J., Ni{\v{c}}kovi{\'{c}}, D.
  (eds.) Runtime Verification. pp. 383--403. Springer International Publishing,
  Cham (2020)

\bibitem{DBLP:conf/cpp/ConradTGPD22}
Conrad, E., Titolo, L., Giannakopoulou, D., Pressburger, T., Dutle, A.: A
  compositional proof framework for {FRET}ish requirements. In: Popescu, A.,
  Zdancewic, S. (eds.) {CPP} '22: 11th {ACM} {SIGPLAN} International Conference
  on Certified Programs and Proofs, Philadelphia, PA, USA, January 17 - 18,
  2022. pp. 68--81. {ACM} (2022). \doi{10.1145/3497775.3503685},
  \url{https://doi.org/10.1145/3497775.3503685}

\bibitem{DBLP:conf/cpp/CoquandS11}
Coquand, T., Siles, V.: {A Decision Procedure for Regular Expression
  Equivalence in Type Theory}. In: Jouannaud, J., Shao, Z. (eds.) CPP. LNCS,
  vol.~7086, pp. 119--134. Springer (2011).
  \doi{10.1007/978-3-642-25379-9\_11},
  \url{https://doi.org/10.1007/978-3-642-25379-9\_11}

\bibitem{DBLP:journals/logcom/Coupet-Grimal03}
Coupet{-}Grimal, S.: An axiomatization of linear temporal logic in the calculus
  of inductive constructions. J. Log. Comput.  \textbf{13}(6),  801--813
  (2003). \doi{10.1093/LOGCOM/13.6.801},
  \url{https://doi.org/10.1093/logcom/13.6.801}

\bibitem{DBR21}
Dabney, J.B., Badger, J.M., Rajagopal, P.: Adding a verification view for an
  autonomous real-time system architecture. In: Proceedings of SciTech Forum.
  p. Online. 2021-0566, {AIAA} (January 2021).
  \doi{https://doi.org/10.2514/6.2021-0566}

\bibitem{DRB22}
Dabney, J.B., Rajagopal, P., Badger, J.M.: Using assume-guarantee contracts for
  developmental verification of autonomous spacecraft. Flight Software Workshop
  (FSW) Online: \url{https://www.youtube.com/watch?v=HFnn6TzblPg} (February
  2022)

\bibitem{DBLP:conf/ifm/ElwingGSTWR23}
Elwing, J., Gamboa{-}Guzman, L., Sorkin, J., Travesset, C., Wang, Z., Rozier,
  K.Y.: Mission-time {LTL} {(MLTL)} formula validation via regular expressions.
  In: Herber, P., Wijs, A. (eds.) iFM. LNCS, vol. 14300, pp. 279--301. Springer
  (2023). \doi{10.1007/978-3-031-47705-8\_15},
  \url{https://doi.org/10.1007/978-3-031-47705-8\_15}

\bibitem{CAVA_LTL_Modelchecker-AFP}
Esparza, J., Lammich, P., Neumann, R., Nipkow, T., Schimpf, A., Smaus, J.G.: A
  fully verified executable {LTL} model checker. Archive of Formal Proofs  (May
  2014), \url{https://isa-afp.org/entries/CAVA\_LTL\_Modelchecker.html}, Formal
  proof development

\bibitem{GMRPSS20}
Giannakopoulou, D., Mavridou, A., Rhein, J., Pressburger, T., Schumann, J.,
  Shi, N.: Formal requirements elicitation with {FRET}. In: International
  Working Conference on Requirements Engineering: Foundation for Software
  Quality (REFSQ-2020). No. ARC-E-DAA-TN77785 (2020)

\bibitem{DBLP:phd/de/Haftmann2009}
Haftmann, F.: {Code generation from specifications in higher-order logic}.
  Ph.D. thesis, Technical University Munich (2009),
  \url{http://mediatum2.ub.tum.de/node?id=886023}

\bibitem{HJRW23}
Hariharan, G., Jones, P.H., Rozier, K.Y., Wongpiromsarn, T.: Maximum
  satisfiability of {M}ission-time {Linear Temporal Logic}. In: Petrucci, L.,
  Sproston, J. (eds.) FORMATS. LNCS, vol. 14138, pp. 86--104. Springer (2023).
  \doi{10.1007/978-3-031-42626-1\_6},
  \url{https://doi.org/10.1007/978-3-031-42626-1\_6}

\bibitem{DBLP:conf/esop/HupelN18}
Hupel, L., Nipkow, T.: {A Verified Compiler from Isabelle/HOL to CakeML}. In:
  Ahmed, A. (ed.) ESOP. LNCS, vol. 10801, pp. 999--1026. Springer (2018).
  \doi{10.1007/978-3-319-89884-1\_35},
  \url{https://doi.org/10.1007/978-3-319-89884-1\_35}

\bibitem{JJKRZ23}
Johannsen, C., Jones, P., Kempa, B., Rozier, K.Y., Zhang, P.: {R2U2 Version
  3.0: Re-Imagining a Toolchain for Specification, Resource Estimation, and
  Optimized Observer Generation for Runtime Verification in Hardware and
  Software}. In: Enea, C., Lal, A. (eds.) Computer Aided Verification. pp.
  483--497. Springer Nature Switzerland, Cham (2023)

\bibitem{KZJZR20}
Kempa, B., Zhang, P., Jones, P.H., Zambreno, J., Rozier, K.Y.: {Embedding
  Online Runtime Verification for Fault Disambiguation on Robonaut2}. In:
  FORMATS. pp. 196--214. LNCS, Springer, Vienna, Austria (September 2020),
  \url{http://research.temporallogic.org/papers/KZJZR20.pdf}

\bibitem{nuXmv-v1.1.0}
Kessler, F.B.: {nuXmv 1.1.0 (2016-05-10) Release Notes}.
  \url{https://es-static.fbk.eu/tools/nuxmv/downloads/NEWS.txt} (2016)

\bibitem{Mission_Time_LTL-AFP}
Kosaian, K., Wang, Z., Sloan, E.: Mission-time linear temporal logic. Archive
  of Formal Proofs  (January 2025),
  \url{https://isa-afp.org/entries/Mission_Time_LTL.html}, Formal proof
  development

\bibitem{KWSR24}
Kosaian, K., Wang, Z., Sloan, E., Rozier, K.: Formalizing {MLTL} formula
  progression in {Isabelle/HOL} (2024), \url{https://arxiv.org/abs/2410.03465}

\bibitem{KraussThesis}
Krauss, A.: Automating Recursive Definitions and Termination Proofs in
  Higher-Order Logic. Ph.D. thesis, Technische Universität München (2009)

\bibitem{DBLP:journals/jar/KraussN12}
Krauss, A., Nipkow, T.: {Proof Pearl: Regular Expression Equivalence and
  Relation Algebra}. J. Autom. Reason.  \textbf{49}(1),  95--106 (2012).
  \doi{10.1007/S10817-011-9223-4},
  \url{https://doi.org/10.1007/s10817-011-9223-4}

\bibitem{LVR19}
Li, J., Vardi, M.Y., Rozier, K.Y.: {Satisfiability Checking for Mission-Time
  LTL}. In: {Proceedings of 31st International Conference on Computer Aided
  Verification (CAV 2019)}. {LNCS}, Springer, New York, NY, USA (July 2019)

\bibitem{LJBHCLR22}
Luppen, Z., Jacks, M., Baughman, N., Hertz, B., Cutler, J., Lee, D.Y., Rozier,
  K.Y.: {Elucidation and Analysis of Specification Patterns in Aerospace System
  Telemetry}. In: {Proceedings of the 14th NASA Formal Methods Symposium (NFM
  2022)}. Lecture Notes in Computer Science (LNCS), vol. 13260. Springer, Cham,
  Caltech, California, USA (May 2022). \doi{10.1007/978-3-031-06773-0\_28}

\bibitem{MN04}
Maler, O., Nickovic, D.: Monitoring temporal properties of continuous signals.
  In: Formal Techniques, Modelling and Analysis of Timed and Fault-Tolerant
  Systems, pp. 152--166. Springer (2004)

\bibitem{GHC}
Marlow, S., Jones, S.L.P.: The {G}lasgow {H}askell {C}ompiler (2012),
  \url{https://api.semanticscholar.org/CorpusID:35370}

\bibitem{NASA-FRET}
{NASA Technology Transfer Program}: {FRET : Formal Requirements Elicitation
  Tool (ARC-18066-1)}. Online:
  \url{https://software.nasa.gov/software/ARC-18066-1} (2024)

\bibitem{Per23}
Perez, I.: Runtime verification with ogma. In: Invited Talk to University of
  California (2023)

\bibitem{ogma}
Perez, I., Goodloe, A.: {OGMA}. \url{https://github.com/nasa/ogma} (2021)

\bibitem{PMPGG22}
Perez, I., Mavridou, A., Pressburger, T., Goodloe, A., Giannakopoulou, D.:
  Automated translation of natural language requirements to runtime monitors.
  In: Fisman, D., Rosu, G. (eds.) Tools and Algorithms for the Construction and
  Analysis of Systems. pp. 387--395. Springer International Publishing, Cham
  (2022)

\bibitem{DBLP:conf/birthday/PnueliA03}
Pnueli, A., Arons, T.: {TLPVS:} {A} {PVS}-based {LTL} verification system. In:
  Dershowitz, N. (ed.) Verification: Theory and Practice, Essays Dedicated to
  Zohar Manna on the Occasion of His 64th Birthday. LNCS, vol.~2772, pp.
  598--625. Springer (2003). \doi{10.1007/978-3-540-39910-0\_26},
  \url{https://doi.org/10.1007/978-3-540-39910-0\_26}

\bibitem{raszyk2020multi}
Raszyk, M., Basin, D., Traytel, D.: Multi-head monitoring of metric dynamic
  logic. In: International Symposium on Automated Technology for Verification
  and Analysis. pp. 233--250. Springer (2020)

\bibitem{RRS14}
Reinbacher, T., Rozier, K.Y., Schumann, J.: Temporal-logic based runtime
  observer pairs for system health management of real-time systems. In:
  Proceedings of the 20th International Conference on Tools and Algorithms for
  the Construction and Analysis of Systems (TACAS). Lecture Notes in Computer
  Science (LNCS), vol.~8413, pp. 357--372. Springer-Verlag (April 2014)

\bibitem{DBLP:conf/vmcai/Roohi018}
Roohi, N., Viswanathan, M.: Revisiting {MITL} to fix decision procedures. In:
  Dillig, I., Palsberg, J. (eds.) VMCAI. LNCS, vol. 10747, pp. 474--494.
  Springer (2018). \doi{10.1007/978-3-319-73721-8\_22},
  \url{https://doi.org/10.1007/978-3-319-73721-8\_22}

\bibitem{Roz16}
Rozier, K.Y.: Specification: The biggest bottleneck in formal methods and
  autonomy. In: {Proceedings of 8th Working Conference on Verified Software:
  Theories, Tools, and Experiments (VSTTE 2016)}. {LNCS}, vol.~9971, pp. 1--19.
  Springer-Verlag, Toronto, ON, Canada (July 2016).
  \doi{10.1007/978-3-319-48869-1\_2}

\bibitem{RS17}
Rozier, K.Y., Schumann, J.: {R2U2: Tool Overview}. In: {Proceedings of
  International Workshop on Competitions, Usability, Benchmarks, Evaluation,
  and Standardisation for Runtime Verification Tools (RV-CUBES)}. vol.~3, pp.
  138--156. Kalpa Publications, Seattle, WA, USA (September 2017),
  \url{https://easychair.org/publications/paper/Vncw}

\bibitem{LTL_to_GBA-AFP}
Schimpf, A., Lammich, P.: Converting linear-time temporal logic to generalized
  {B}üchi automata. Archive of Formal Proofs  (May 2014),
  \url{https://isa-afp.org/entries/LTL\_to\_GBA.html}, Formal proof development

\bibitem{LTL_Master_Theorem-AFP}
Seidl, B., Sickert, S.: A compositional and unified translation of {LTL} into
  $\omega$-automata. Archive of Formal Proofs  (April 2019),
  \url{https://isa-afp.org/entries/LTL\_Master\_Theorem.html}, Formal proof
  development

\bibitem{LTL_to_DRA-AFP}
Sickert, S.: Converting linear temporal logic to deterministic (generalized)
  {R}abin automata. Archive of Formal Proofs  (September 2015),
  \url{https://isa-afp.org/entries/LTL\_to\_DRA.html}, Formal proof development

\bibitem{LTL-AFP}
Sickert, S.: Linear temporal logic. Archive of Formal Proofs  (March 2016),
  \url{https://isa-afp.org/entries/LTL.html}, Formal proof development

\bibitem{LTL_Normal_Form-AFP}
Sickert, S.: An efficient normalisation procedure for linear temporal logic:
  {Isabelle/HOL} formalisation. Archive of Formal Proofs  (May 2020),
  \url{https://isa-afp.org/entries/LTL\_Normal\_Form.html}, Formal proof
  development

\bibitem{LL1ParserAFP}
Tilscher, S., Wimmer, S.: {LL}(1) parser generator. Archive of Formal Proofs
  (May 2024), \url{https://isa-afp.org/entries/LL1\_Parser.html}, formal proof
  development

\bibitem{fret-proof-framework}
Titolo, L., Conrad, E., Giannakopoulou, D., Pressburger, T., Dutle, A.: {FRET
  Proof Framework}.
  \url{https://lauratitolo.github.io/project/fret-proof-framework/} (2022)

\bibitem{WEST2024tool}
Wang, Z., Gamboa-Guzman, L.P., Rozier, K.Y.: {WEST: Interactive Validation of
  Mission-time Linear Temporal Logic (MLTL)} (2024),
  \url{https://temporallogic.org/research/WEST/}

\bibitem{Mission_Time_LTL_to_Regular_Expression-AFP}
Wang, Z., Kosaian, K.: Mission-time linear temporal logic to regular
  expressions. Archive of Formal Proofs  (January 2025),
  \url{https://isa-afp.org/entries/Mission_Time_LTL_to_Regular_Expression.html},
  Formal proof development

\bibitem{MyhillNerodeFormalization}
Wu, C., Zhang, X., Urban, C.: A formalisation of the {M}yhill-{N}erode theorem
  based on regular expressions (proof pearl). In: van Eekelen, M., Geuvers, H.,
  Schmaltz, J., Wiedijk, F. (eds.) Interactive Theorem Proving. pp. 341--356.
  Springer Berlin Heidelberg, Berlin, Heidelberg (2011)

\bibitem{ZADJR23}
Zhang, P., Aurandt, A.A., Dureja, R., Jones, P.H., Rozier, K.Y.: Model
  predictive runtime verification for cyber-physical systems with real-time
  deadlines. In: Petrucci, L., Sproston, J. (eds.) Formal Modeling and Analysis
  of Timed Systems - 21st International Conference, {FORMATS} 2023, Antwerp,
  Belgium, September 19-21, 2023, Proceedings. Lecture Notes in Computer
  Science, vol. 14138, pp. 158--180. Springer (2023).
  \doi{10.1007/978-3-031-42626-1\_10},
  \url{https://doi.org/10.1007/978-3-031-42626-1\_10}

\bibitem{Lean_Extended_regex}
Zhuchko, E., Veanes, M., Ebner, G.: Lean formalization of extended regular
  expression matching with lookarounds. In: Proceedings of the 13th ACM SIGPLAN
  International Conference on Certified Programs and Proofs. p. 118–131. CPP
  2024, Association for Computing Machinery, New York, NY, USA (2024).
  \doi{10.1145/3636501.3636959}, \url{https://doi.org/10.1145/3636501.3636959}

\end{thebibliography}

\appendix

\newpage
\section{Experiments}\label{app:experiments}
The experiments presented in the main body of the paper (see \rref{fig:cactus}) aggregate random formulas of varying complexity into one single plot, demonstrating the overall performance of the three WEST implementations relative to each other on a sizable random benchmark set.
That is, for set values of the parameters (number of atomic propositions $n$, maximum interval time bound $b$, and maximum nesting depth $d$), we generate formulas with complexity \textit{at most} nesting depth $d$ and time bound $b$.

\begin{figure}[!htb]
\begin{minipage}{0.5\textwidth}
    \centering
    \includegraphics[width=1\linewidth]{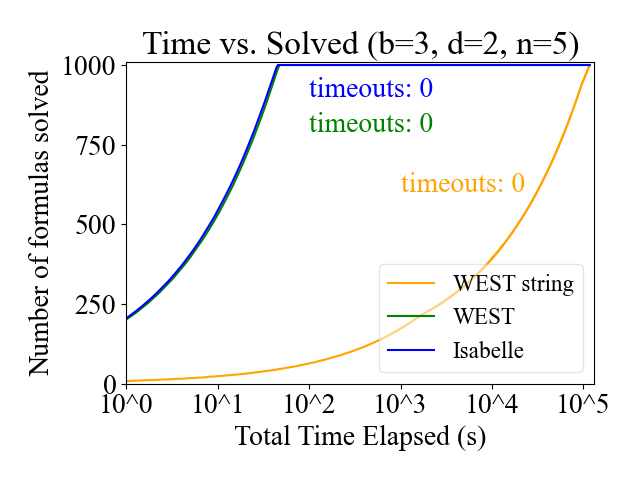}
\end{minipage}
\hfill
\begin{minipage}{0.5\textwidth}
    \centering
    \includegraphics[width=1\linewidth]{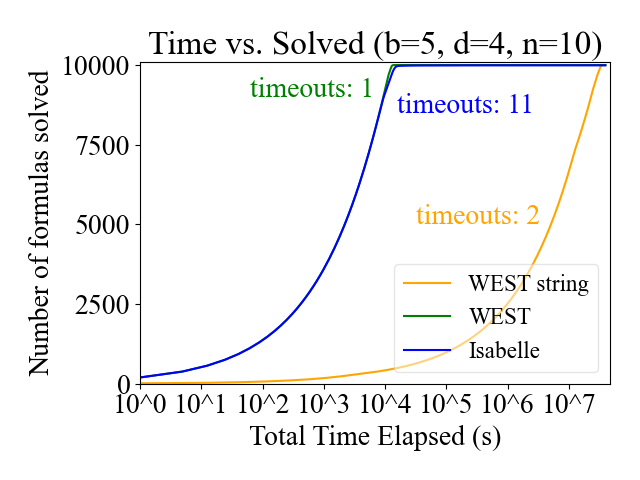}
\end{minipage}
\caption{Aggregate cactus plots comparing the three WEST implementations on relatively simple formulas.
On the left, with formulas of max depth $d=2$, max time bound $b=3$, and $n=5$ atomic propositions, the Isabelle implementation and optimized WEST are nearly identical in performance.
On the right, with formulas of max depth $d=4$, max time bound $b=5$, and $n=10$ atomic propositions, the Isabelle implementation is overall extremely close in performance to the optimized WEST implementation, but times out on a few more formulas than both unverified WEST implementations.
}
\label{fig:experiments_easy2cactus}
\end{figure}

It took some time to identify interesting combinations of these parameters to consider.
When we generate easier formulas with lower nesting depth and time bounds, we find that the Isabelle implementation is nearly identical in performance to optimized WEST (see \rref{fig:experiments_easy2cactus}).

When determining which parameters would be interesting choices for the aggregate plots, we found it beneficial to run some more granular experiments on smaller batches of formulas. 
We present some of those results now, where we separate out runtimes by individual values of the parameters $b$, and $d$. 
We also include the number of timeouts for each implementation in each batch in separate plots.

\begin{figure}[!htb]
\begin{minipage}{0.48\textwidth}
    \centering
    \includegraphics[width=1\linewidth]{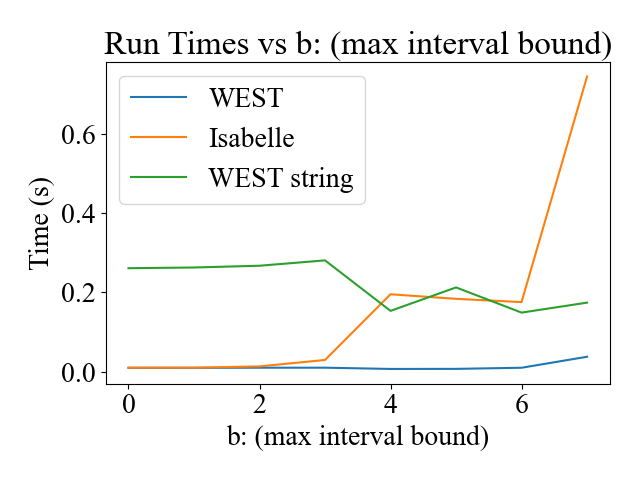}
\end{minipage}
\hfill
\begin{minipage}{0.48\textwidth}
    \centering
    \includegraphics[width=1\linewidth]{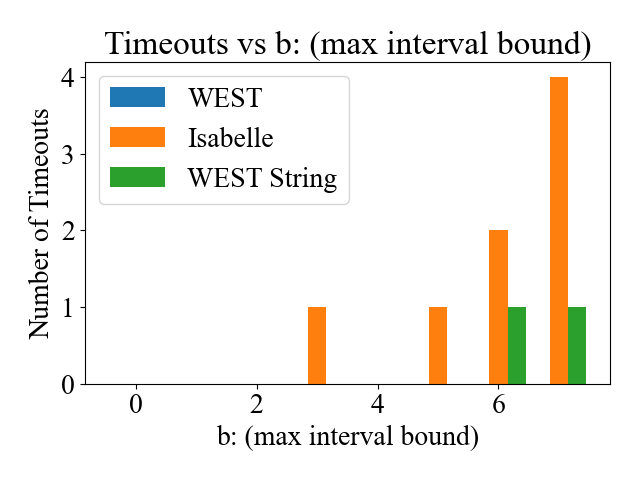}
\end{minipage}
\caption{Results for $n=5$, $d=2$, and values of $b$ from $0$ to $7$, with a batch size of 300 formulas per value of $b$.
We see that the Isabelle implementation times out slightly more than the other tools (8 timeouts total, compared to 2 for unoptimized WEST and 0 for optimized WEST, out of the 2400 examples total).
}
\label{fig:experiments_easyb}
\end{figure}

\begin{figure}[!htb]
\begin{minipage}{0.48\textwidth}
    \centering
    \includegraphics[width=1\linewidth]{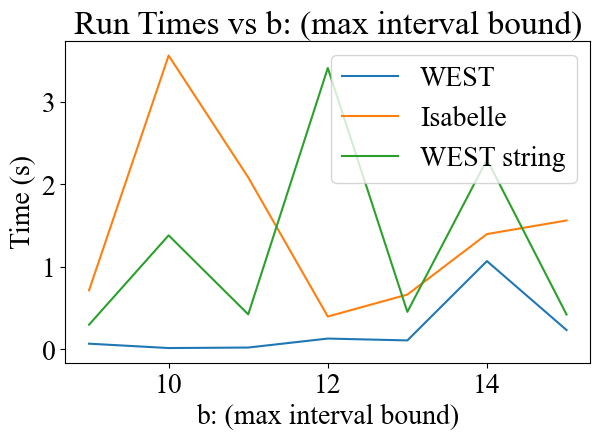}
\end{minipage}
\hfill
\begin{minipage}{0.48\textwidth}
    \centering
    \includegraphics[width=1\linewidth]{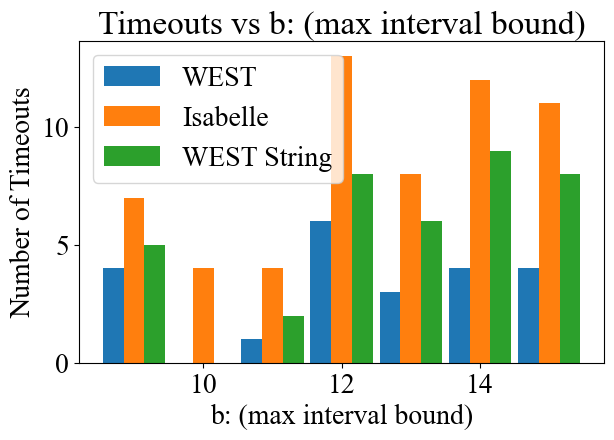}
\end{minipage}
\caption{Results for $n=5$, $d=4$, and varying values of $b$ from $8$ to $15$, with a batch size of 50 formulas per value of $b$.
Here, both the Isabelle implementation and the unoptimized version of WEST exhibit some slightly more ``spiky'' behavior than optimized WEST. The Isabelle implementation also times out slightly more overall than the unoptimized tools.
}
\label{fig:experiments_hardb}
\end{figure}

\begin{figure}[!htb]
\begin{minipage}{0.48\textwidth}
    \centering
    \includegraphics[width=1\linewidth]{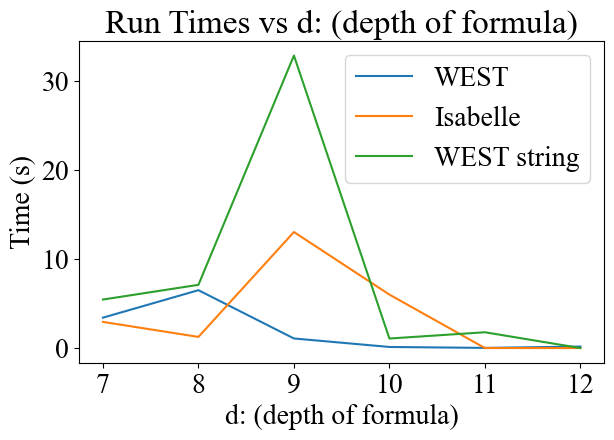}
\end{minipage}
\hfill
\begin{minipage}{0.48\textwidth}
    \centering
    \includegraphics[width=1\linewidth]{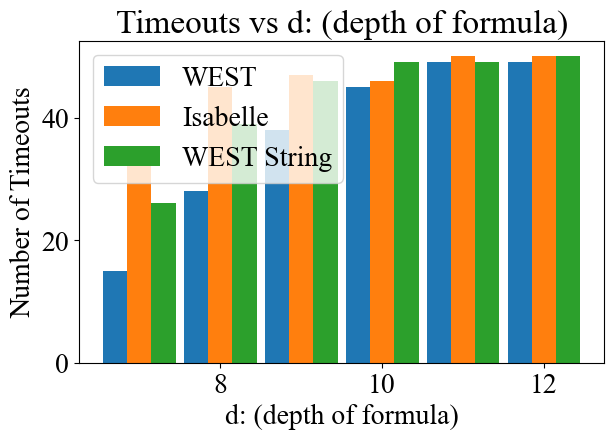}
\end{minipage}
\caption{Results for $n=5$, $b=3$, and varying values of $d$ from $7$ to $12$, with a batch size of 50 formulas per value of $d$. 
These values of $d$ are challenging for the tools with our timeout of 60 seconds.
The Isabelle implementation times out on most of the formulas; the optimized WEST tool steadily increases in timeouts across increasing values of $d$. 
}
\label{fig:experiments_hardd}
\end{figure}

In each experiment, we vary one parameter (either the depth of the formula $d$ or the maximum time bound of the formula $b$) over a range of values, while keeping the other parameters fixed, and generate a batch of random formulas for each value of the varying parameter; the batch sizes vary by experiment (for more complicated formulas, we use smaller batches).
Then we run the three implementations on each formula in the batch, recording the average runtime of each batch.
So as not to skew the averages, we do not include timeouts in this average runtime; instead, we record the number of timeouts in separate plots.
Each experiment had a timeout of 60 seconds.
In these experiments, the highly optimized version of WEST is generally fastest, as expected.
The Isabelle code and the unoptimized version of WEST exhibit some ``spiky'' behavior, where there is a little more variability---some batch sizes or depths take longer than others (i.e., the results are nonlinear). 
This is likely influenced by our choice to separate timeouts from the average runtime and also possibly due to certain formula shapes being especially difficult.
We present the results of experiments on varying values of $b$ in \rref{fig:experiments_easyb} and \rref{fig:experiments_hardb}, and on varying higher values of $d$ in \rref{fig:experiments_hardd} (which is a continuation of the results in \rref{fig:experiments_easyd}).

\begin{figure}[!htb]
\begin{minipage}{0.48\textwidth}
    \centering
    \includegraphics[width=1\linewidth]{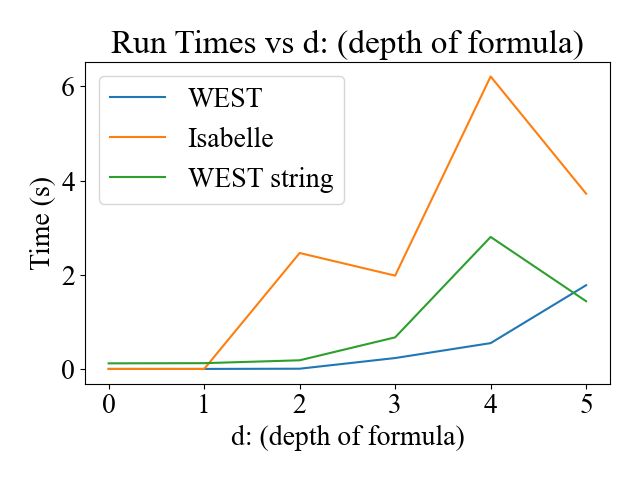}
\end{minipage}
\hfill
\begin{minipage}{0.48\textwidth}
    \centering
    \includegraphics[width=1\linewidth]{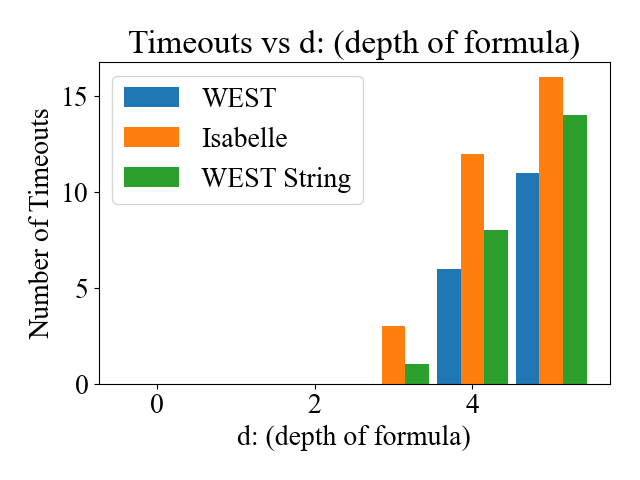}
\end{minipage}
\caption{Results for $n=5$, $b=3$, and varying values of $d$ from $0$ to $5$, with a batch size of 25 formulas per value of $d$.
Each formula in this experiment is generated with nested release and until operators.
The Isabelle implementation is, on average, slightly slower than the unverified implementations and times out on more formulas (note that, in the last data point where $d=5$, the average runtimes of the Isabelle implementation and the unoptimized WEST tool decrease because of the numerous timeouts, which are not included in the average timing information).
}
\label{fig:experiments_nestedUR}
\end{figure}

Lastly, we perform an experiment specifically on formulas with nested release and until operators, varying the maximum nesting depth $d$.
Such formulas are considered in the runtime analysis of the WEST tool \cite{DBLP:conf/ifm/ElwingGSTWR23} because this formula shape leads to the worst theoretical runtime for the WEST algorithm.
We present the results of this experiment in \rref{fig:experiments_nestedUR}.

\end{document}